\documentclass[aps,twocolumn,showpacs]{revtex4}

\usepackage{graphicx,epsfig,verbatim,enumerate}
\usepackage{amsmath,amssymb}
\usepackage{ifthen}

\newcommand{\thnote}[1]{}

\newcommand{\etal}{\textit{et al.}}
\newcommand{\www}[1]{\url{#1}}
\newcommand{\req}[1]{(\ref{#1})}

\newcommand{\Om}{\Omega}
\newcommand{\om}{\omega}

\newcommand{\tdiff}[2]{\mbox{d} #1/\mbox{d} #2}

\newcommand{\rbfrac}{\genfrac(){}0}

\newcommand{\cprob}[2]{P(#1\,|\,#2)}

\newcommand{\dee}[1]{\mbox{d}#1}

\newcommand{\Tmunu}{T_{\mu,\nu}}

\newcommand{\avg}[1]{\left\langle#1\right\rangle}
\newcommand{\tavg}[1]{\langle#1\rangle}

\newcommand{\okell}{{l^{\mathrm{(s)}}}}

\newcommand{\okellb}{l_{\mu,\nu}^{\mathrm{(s,\, b)}}}
\newcommand{\okelli}{l_{\mu,\nu}^{\mathrm{(s,\, i)}}}
\newcommand{\okellinum}[1]{l_{\mu,\nu=#1}^{\mathrm{(s,\, i)}}}
\newcommand{\okelle}{l_{\mu,\nu}^{\mathrm{s,\, e}}}

\newcommand{\okellsnum}[1]{l_{#1}^{\mathrm{\, (s)}}}

\begin{document}

\title{
Geometry of River Networks III:\\ Characterization of Component Connectivity
}

\author{
  \firstname{Peter Sheridan}
  \surname{Dodds}
  }
\thanks{Author to whom correspondence should be addressed}
\email{dodds@segovia.mit.edu}
\homepage{http://segovia.mit.edu/}
\affiliation{Department of 
Mathematics and Department of Earth, 
Atmospheric and Planetary Sciences,
Massachusetts Institute of Technology,
Cambridge, MA 02139.}

\author{
  \firstname{Daniel H.}
  \surname{Rothman}
  }
\email{dan@segovia.mit.edu}
\affiliation{Department  of Earth, 
Atmospheric and Planetary Sciences,
Massachusetts Institute of Technology, 
Cambridge, MA 02139.}

\date{\today}

\begin{abstract}
River networks serve as a paradigmatic example of all branching
networks.  Essential to understanding the overall structure of river
networks is a knowledge of their detailed architecture.  Here we show
that sub-branches are distributed exponentially in size and that they
are randomly distributed in space, thereby completely characterizing
the most basic level of river network description.  Specifically, an
averaged view of network architecture is first provided by a proposed
self-similarity statement about the scaling of drainage density, a
local measure of stream concentration.  This scaling of drainage
density  is shown to imply Tokunaga's law, a description of the
scaling of  side branch abundance along a given stream, as well as a
scaling law for stream lengths.  This establishes the scaling of the
length scale associated with drainage density as the basic signature
of self-similarity in river networks.  We then consider fluctuations
in drainage density and consequently the numbers of side branches.
Data is analyzed for the Mississippi River basin  and a model of
random directed networks.  Numbers of side streams are found to follow
exponential distributions as are stream lengths and inter-tributary
distances along streams.  Finally, we derive the joint variation  of
side stream abundance with stream length, affording a full description
of fluctuations in network structure.  Fluctuations in side stream
numbers are shown to be a direct result of fluctuations in stream
lengths.  This is the last paper in a series of three on the geometry
of river networks.
\end{abstract}

\maketitle

\section{Introduction}
\label{sec:tokunaga.intro}

This is the last paper in a series of three on the geometry
of river networks.
In the first~\cite{dodds2000ua} we examine
in detail the description of river networks by
scaling laws~\cite{maritan96a,rodriguez-iturbe97,dodds99pa,dodds2000pa}
and the evidence for universality.
Additional introductory remarks concerning the motivation
of the overall work are to found in this first paper.
In the second article~\cite{dodds2000ub} we address
distributions of the basic components of river networks, stream
segments and sub-networks.  
Here, we provide an analysis complementary to
the work of the second paper
by establishing a description of how river network
components fit together.
As before, we are motivated by the premise that
while relationships of mean quantities are primary in
any investigation, the behavior of higher order moments potentially
and often do encode significant information.

Our purpose then is to investigate the 
distributions of quantities which describe
the architecture of river networks.  
The goal is to quantify these distributions and, where this is
not possible, to quantify fluctuations.
In particular, we center
our attention on Tokunaga's law~\cite{tokunaga66,tokunaga78,tokunaga84}
which is a statement about network
architecture describing the tributary structure of streams.  
Since Tokunaga's law
can be seen as the main part of a platform from which
all other river network scaling laws follow~\cite{dodds99pa},
it is an obvious starting point for the investigation
of fluctuations in river network structure.
We use data from 
the Schediegger model of random networks~\cite{scheidegger67} 
and the Mississippi river.
We find the distributions obtained from these
two disparate sources agree very well in form.
We are able to write down scaling forms of
all distributions studied.
We observe a number of distributions
to be exponential, therefore requiring only
one parameter for their description.
As a result, we introduce a dimensionless scale $\xi_t$,
finding it to be sufficient
to describe the fluctuations present
in Tokunaga's law and thus
potentially all river network scaling laws.
Significantly, we observe the spatial distribution
of stream segments to be random implying we
have reached the most basic description
of network architecture.

Tokunaga's law is also intimately connected with
drainage density, $\rho$, a quantity which
will be used throughout the paper.
Drainage density is a measure
of stream concentration or, equivalently,
how a network fills space.
We explore this connection in detail, showing
how simple assumptions regarding drainage
density lead to Tokunaga's law.

The paper is structured as follows.
We first outline Horton-Strahler stream ordering
which provides the necessary descriptive taxonomy
for river network architecture.
We then define Tokunaga's law and 
introduce a scaling law for a specific form of drainage density.
We briefly describe Horton's laws for stream number and length
and some simple variations. (Both stream ordering
and Horton's laws are covered in more detail in~\cite{dodds2000ub}).
We show that the scaling law of drainage density may
be taken as an assumption from which all other scaling
laws follow.
We also briefly consider
the variation of basin shapes (basin allometry)
in the context of directedness.
This brings us to the focal point of the paper, the identification
of a statistical generalization of Tokunaga's law.
We first examine distributions of numbers of
tributaries (side streams) and
compare these with distributions of stream segment lengths.
We observe both distributions to be exponential leading
to the notion that stream segments are distributed
randomly throughout a network.
The presence of exponential distributions
also leads to the introduction of
the characteristic number $\xi_t$ and
the single length-scale $\xi_\okell \propto \xi_t$.
We then study the variation of
tributary spacing along streams so as to understand fluctuations
in drainage density and again find the signature of randomness.
This leads us to develop a joint probability distribution
connecting the length of a stream
with the frequency of its side streams.

\section{Definitions}
\subsection{Stream ordering}
\label{sec:tokunaga.streamordering}
Horton-Strahler stream ordering\cite{horton45,strahler57} 
breaks a river network down
into a set of \textit{stream segments}.
The method can be thought of as an iterative pruning.
First, we define a \textit{source stream} as the
stream section that runs from a channel head
to the first junction with another stream.
These source streams are classified as the
\textit{first-order} stream segments of the network.
Next, we remove all
source streams and 
identify the new source streams of the remaining network.
These are the network's \textit{second-order} stream segments.  
The process is repeated until one stream segment is left of order $\Omega$.
The order of the basin is then defined to be $\Omega$
(we will use the words basin and network interchangeably).

In discussing network architecture,
we will speak of \textit{side streams} and 
\textit{absorbing streams}.  A side stream is any stream
that joins into a stream of higher order,
the latter being the absorbing stream.  
We will denote the orders of absorbing and side streams
by $\mu$ and $\nu$ but
when referring to an isolated stream or streams where
their relative rank is ambiguous,
we will write stream order as $\om$, subscripted
as seems appealing.

Central to our investigation of network architecture
is stream segment length.  
As in~\cite{dodds2000ub}, we denote
this length by $\okellsnum{\om}$ for a stream segment
of order $\om$.  We will also introduce a number
of closely related lengths which describe
distances between side streams.
When referring
to streams throughout we will specifically mean stream
segments of a particular order unless otherwise indicated.
This is to avoid confusion with
the natural definition of a stream which is
the path from a point on a network moving 
upstream to the most distant source.
For an order $\om$ basin,
we denote this \textit{main stream length} by $l_\om$

Note also that we consider river networks in planform, i.e.,
as networks projected onto the horizontal 
(or gravitationally flat) plane.
This simplification poses no great 
concern for the analysis of large scale networks such as the Mississippi
but must be considered in the context of drainage basins
with significant relief.

\subsection{Tokunaga's law}
\label{sec:tokunaga.tokslaw}

\begin{figure}[tb!]
  \begin{center}
    \ifthenelse{\boolean{@twocolumn}}
    {
      \epsfig{file=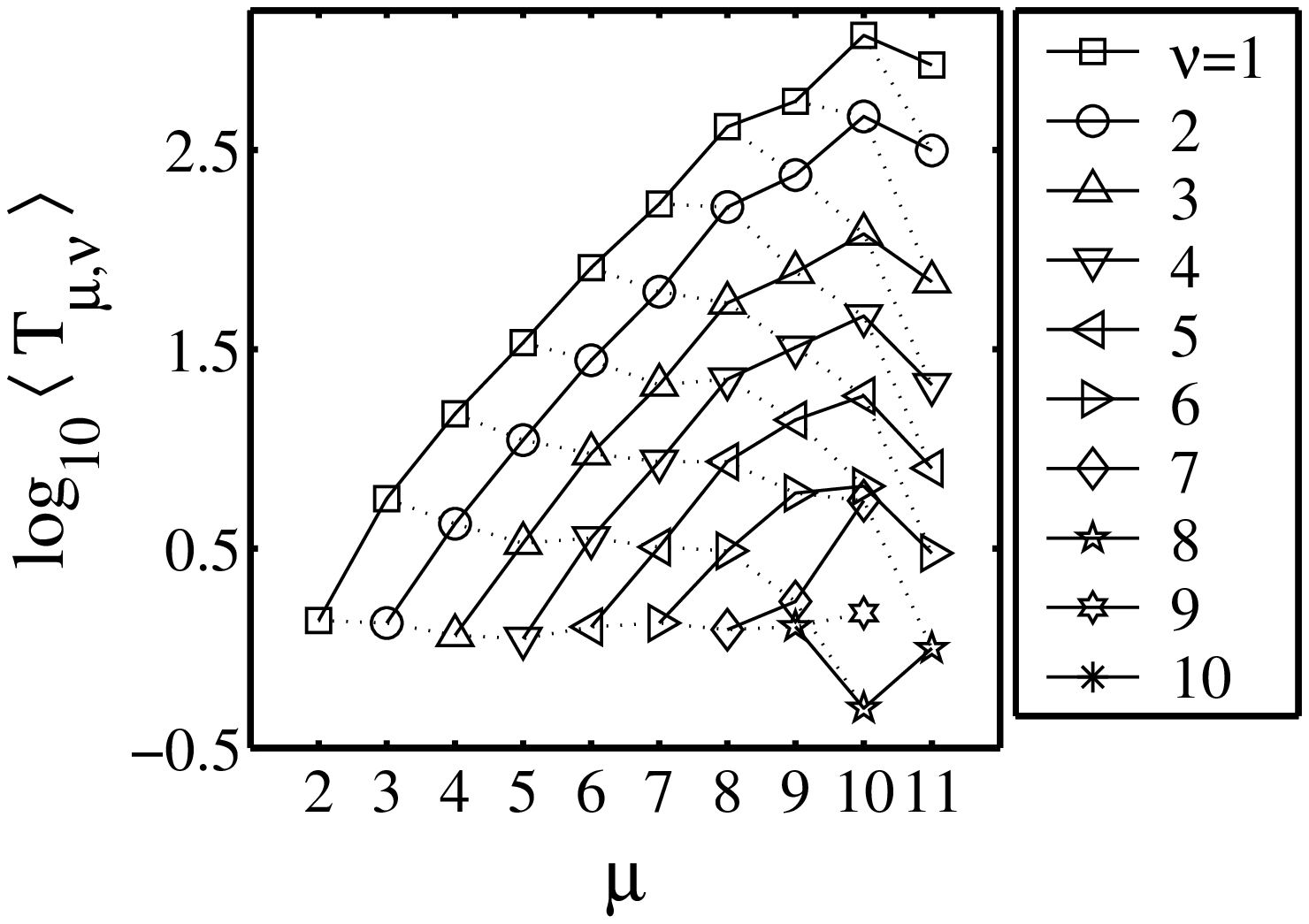,width=0.48\textwidth}
      }
    {
      \epsfig{file=figtok_mispi10_3_noname.ps,width=0.64\textwidth}
      }
    \caption[Tokunaga graph for the Mississippi]{
      A Tokunaga graph for the Mississippi River.
      The values are given in 
      Table~\protect\ref{tab:tokunaga.mispitokratios}.
      Each point represents a Tokunaga ratio $\tavg{\Tmunu}$.
      The solid lines follow variations in the order of the absorbing
      stream $\mu$ while the dotted lines follow unit increments
      in both $\mu$ and $\nu$, the order of side streams.
      In comparison,
      the Tokunaga graph of an exactly self-similar network 
      would have points evenly spaced at $\ln{T_1} + (\mu+\nu-1)\ln{R_T}$
      where $1 \le \nu < \mu = 2,3,\ldots,\Om$, i.e.,
      all lines in the plot would be straight and uniformly spaced,
      with the dotted lines being horizontal.
      The nature of deviations in scalings laws for river
      networks is addressed in~\cite{dodds2000ua}.
      }
    \label{fig:tokunaga.tok_mispi10}
  \end{center}
\end{figure}

Defining a stream ordering on a network allows 
for a number of well-defined measures of connectivity,
stream lengths and drainage areas.  
Around a decade after the Strahler-improved\nocite{strahler57}
stream ordering of Horton\nocite{horton45} appeared, 
Tokunaga\nocite{tokunaga66}
introduced the idea of measuring side 
stream statistics~\cite{tokunaga66,tokunaga78,tokunaga84}.
This technique arguably provides
the most useful measurement based
on stream ordering but has only recently received
much attention~\cite{dodds99pa,cui99,turcotte98,peckham95}.
The idea is simply, for a given network, to count the average number of
order $\nu$ side streams entering an order $\mu$
absorbing stream.  This gives $\tavg{\Tmunu}$, a set of double-indexed
parameters for a basin.
Note that $\Om \ge \mu > \nu \ge 1$,
so we can view the Tokunaga ratios as a lower triangular
matrix.  
An example for the Mississippi river 
is shown in Table~\ref{tab:tokunaga.mispitokratios}~\footnote{
  The network for the Mississippi
  was extracted from a topographic dataset
  constructed from three arc second
  USGS Digital Elevation Maps,
  decimated by averaging to 
  approximately 1000 meter horizontal resolution (\protect\www{www.usgs.gov}).
  At this grid scale, the Mississippi was found to be
  an order $\Om=11$ basin.
  }
The same data is represented pictorially 
in Figure~\ref{fig:tokunaga.tok_mispi10} in
what we refer to as a \textit{Tokunaga graph}.

\begin{table}[tbh!]
  \begin{center}
    \small
    \begin{tabular}{c|cccccccccc} \toprule
      & $\nu=1$ & 2 & 3 & 4 & 5 & 6 & 7 & 8 & 9 & 10 \\ \colrule
      $\mu=2$ & 1.7 & \mbox{} &  \mbox{} &  \mbox{} &  \mbox{} &  \mbox{} &  \mbox{} &  \mbox{} &  \mbox{} & \mbox{}  \\ 
      3 & 4.9 & 1.3  &  \mbox{} &  \mbox{} &  \mbox{} &  \mbox{} &  \mbox{} &  \mbox{} &  \mbox{} &  \mbox{} \\
      4 & 12 & 3.8  & 1.1  &  \mbox{} &  \mbox{} &  \mbox{} &  \mbox{} &  \mbox{} &  \mbox{} &  \mbox{} \\
      5 & 29 & 9.1  & 2.9  & 1.0  &  \mbox{} &  \mbox{} &  \mbox{} &  \mbox{} &  \mbox{} &  \mbox{} \\
      6 & 71 & 23  & 7.7  & 3.0  & 1.2  &  \mbox{} &  \mbox{} &  \mbox{} &  \mbox{} &  \mbox{} \\
      7 & 190 & 56  & 19  & 7.8  & 3.3  & 1.1  &  \mbox{} &  \mbox{} &  \mbox{} &  \mbox{} \\
      8 & 380 & 110  & 39  & 17  & 6.9  & 2.6  & 1.0  &  \mbox{} &  \mbox{} &  \mbox{} \\
      9 & 630 & 170  & 64  & 28  & 11  &  4.5  &    3.0  &  0.60  &  \mbox{} &  \mbox{} \\
      10 & 1100 & 270  &   66  & 29  & 13  & 4.3  & 2.7  &    1  &    1  &  \mbox{} \\
      11 & 1400 &  510  &  120  &   66  &   25  &   12  &    9  &    3  &    1  &    1 \\ \botrule
    \end{tabular}
    \caption[Tokunaga ratios for the Mississippi]{
      Tokunaga ratios for the Mississippi River.
      The row indices are the absorbing stream orders while
      the columns correspond to side stream orders.
      Each entry is the average number of 
      order $\nu$ side streams per order $\mu$ absorbing stream.
      }
    \label{tab:tokunaga.mispitokratios}
  \end{center}
\end{table}

Tokunaga made several key observations about
these side stream ratios.  The first is that because
of the self-similar nature of river networks, the
$\tavg{\Tmunu}$ should not depend absolutely on either of
$\mu$ or $\nu$ but only on the relative difference,
i.e., $k=\mu-\nu$.  The second is that in changing
the value of $k=\mu-\nu$, the $\tavg{\Tmunu}$ must
themselves change by a systematic ratio.
These statements lead to Tokunaga's law:
\begin{equation}
  \label{eq:tokunaga.tokslaw}
  \avg{\Tmunu} = \avg{T_k} = \avg{T_1} (R_T)^{k-1}.
\end{equation}
Thus, only two parameters are necessary to characterize
the set of $\Tmunu$: $T_1$ and $R_T$.

The parameter $T_1 > 0$ is
the average number of side streams of one order
lower than the absorbing stream, typically on the
order of 1.0--1.5.  Since these side streams
of one order less are the dominant side streams
of the basin, their number estimates
the basin's breadth.
In general, larger values of $\tavg{T_1}$ correspond
to wider basins while smaller values are in keeping
with basins with relatively thinner profiles.

The ratio $R_T > 1$ measures how
the density of side streams of decreasing order
increases.  It is a measure of changing
length scales and has a simple interpretation
with respect to Horton's laws which we describe below.
Thus, already inherent in Tokunaga's law
is a generalization of drainage density $\rho$.
the usual definition of which is
given as follows.
For a given 
region of landscape with area $A$
with streams totalling in length $L$, 
$\rho = L/A$ and has the dimensions
of an inverse length scale~\cite{horton45}.
One may think of $\rho$ as the inverse of the typical
distance between streams, i.e., the characteristic
scale beyond which erosion cannot
more finely dissect the landscape~\cite{horton45}.
In principle, drainage density may vary from landscape to landscape
and also throughout a single region.
Below, we will turn this observation about
Tokunaga's law around to show that all river network
scaling laws may be derived from an expanded
notion of drainage density.

Even though the number of side streams
entering any absorbing stream must of course be an integer, 
Tokunaga's ratios are under no similar obligation
since they are averages.
Nevertheless, Tokunaga's law provides a good sense of the structure of
a network albeit at a level of averages.
One of our main objectives here is to go further and consider 
fluctuations about and the full
distributions underlying the $\tavg{\Tmunu}$.

Finally, a third important observation of Tokunaga is that 
two of Horton's laws follow from Tokunaga's law,
which we next discuss.

\subsection{Horton's laws}
\label{sec:tokunaga.hortonslaw}
We review Horton's laws~\cite{horton45,schumm56a,dodds99pa} and then
show how self-similarity and drainage density
lead to Tokunaga's law, Horton's laws and
hence all other river network scaling laws.

The relevant quantities for Horton's relations are $n_\om$,
the number of order $\om$ streams, and 
$\tavg{l_\om}$, the average main stream length (as opposed to
stream segment length $\tavg{\okellsnum{\om}}$) of order $\om$ basins.
The laws are simply that the ratio of these quantities from
order to order remain constant:
\begin{equation}
  \label{eq:tokunaga.hortslaws}
  \frac{n_{\om+1}}{n_\om} = 1/R_n
  \quad
  \mbox{and}
  \quad
  \frac{\avg{l_{\om+1}}}{\avg{l_\om}} = R_l,
\end{equation}
for $\om \ge 1$.
Note the definitions are chosen
so that all ratios are greater than unity.  
The number of streams decreases with 
order while all areas and lengths grow.  

A similar law for basin areas~\cite{horton45,schumm56a}
states that
$\tavg{a_{\om+1}}/\tavg{a_\om} = R_a$
where $\tavg{a_\om}$ is the average drainage area of 
an order $\om$ basin.  However, with the
assumption of uniform drainage density it 
can be shown that $R_n \equiv R_a$~\cite{dodds99pa}
so we are left with the two independent Horton laws
of equation~\req{eq:tokunaga.hortslaws}.

As in~\cite{dodds2000ub}, we consider another Horton-like law for
stream segment lengths:
\begin{equation}
  \label{eq:tokunaga.hortslaws_ell}
  \frac{\avg{\okellsnum{\om+1}}}{\avg{\okellsnum{\om}}} = R_\okell.
\end{equation}
As we will show, the form of 
the distribution of the variable $\Tmunu$ is
a direct consequence of the distribution of $\okellsnum{\om}$.

Tokunaga showed that Horton's laws
of stream number and stream length follow from
what we have called 
Tokunaga's law, equation~\req{eq:tokunaga.tokslaw}.
For example, the solution of a difference 
equation relating the $n_\om$ and the $T_k$
leads to the result
$ R_n =  A_T + \left[A_T^2 - 2R_T\right]^{1/2}$ 
where $A_T = (2 + R_T + T_1)/2$
for $\Om=\infty$ and a more complicated expression
is obtained for finite $\Om$~\cite{dodds99pa,tokunaga78,tokunaga84,peckham95}.
In keeping with our previous remarks on $T_1$,
this expression for $R_n$
shows that an increase in $T_1$ will increase
$R_n$ which, since $R_a \equiv R_n$, corresponds to
a network where basins tend to be relatively broader.
Our considerations will expand significantly
on this connection between the network descriptions of
Horton and Tokunaga.

\section{The implications of a scaling law for drainage density}
\label{sec:tokunaga.dd}

We now introduce a law for drainage density
based on stream ordering.
We write $\rho_{\mu,\nu}$ for the number of 
side streams of order $\nu$ per unit length
of order $\mu$ absorbing stream.
We expect these densities to be 
independent of the order of the absorbing stream
and so we will generally use $\rho_\nu$.
The typical length separating order $\nu$
side streams is then $1/\rho_\nu$.
Assuming self-similarity
of river networks, we must have
\begin{equation}
  \label{eq:tokunaga.hortondd}
  \rho_{\nu+1}/\rho_{\nu} = 1/R_\rho
\end{equation}
where $R_\rho > 1$ independent of $\nu$.

All river network scaling laws in the planform may be seen
to follow from this relationship.
Consider an absorbing
stream of order $\mu$.  Self-similarity immediately
demands that the number of side streams of order $\mu-1$
must be statistically independent of $\mu$.  This number 
is of course $\tavg{T_1}$.  Therefore, the typical length
of an order $\mu$ absorbing stream must be 
\begin{equation}
  \label{eq:tokunaga.tokellconnect}
  \avg{\okellsnum{\mu}} = \avg{T_1}/\rho_{\mu-1}.
\end{equation}
Using equation~\req{eq:tokunaga.hortondd}
to replace $\rho_{\mu-1}$ in the above equation, we find
\begin{equation}
  \label{eq:tokunaga.tokellconnect2}
  T_1/\rho_{\mu-1} = R_\rho T_1/\rho_{\mu-2}.
\end{equation}
Thus, $T_2 = R_\rho T_1$ and, in general
$T_k = (R_\rho)^{k-1} T_1$.  This is Tokunaga's law
and we therefore have 
\begin{equation}
  \label{eq:tokunaga.RrhoRT}
  R_\rho \equiv R_T.
\end{equation}

Equation~\req{eq:tokunaga.hortondd}
and equation~\req{eq:tokunaga.tokellconnect} also give
\begin{equation}
  \label{eq:tokunaga.tokellconnect3}
  \avg{\okellsnum{\mu}} = T_1/\rho_{\mu-1} = R_\rho T_1/\rho_{\mu-2}
  = R_\rho \avg{\okellsnum{\mu-1}}.
\end{equation}
On comparison with equation~\req{eq:tokunaga.hortslaws_ell},
we see that the above
is our Hortonian law of stream segment lengths and that 
\begin{equation}
  \label{eq:tokunaga.RrhoRell}
  R_\rho \equiv R_\okell.  
\end{equation}

As $R_\okell$ is the basic length-scale ratio
in the problem, we rewrite 
equation~\req{eq:tokunaga.hortondd},
our Hortonian law of drainage density, as
\begin{equation}
  \label{eq:tokunaga.hortondd2}
  \rho_{\nu+1}/\rho_{\nu} = 1/R_\okell.
\end{equation}
The above statement 
becomes our definition of the self-similarity
of drainage density.

\section{Basin allometry}
\label{sec:tokunaga.allometry}
Given that we have suggested the need for only a single 
relevant length ratio, we must remark here on basin allometry.
Allometry refers to the relative growth or scaling
of a shape's dimensions and was originally 
introduced in the context of biology~\cite{huxley36}.
A growth or change being allometric usually implies 
it is not self-similar.
A longstanding issue in the study of river networks has
been whether or not basins are 
allometric~\cite{rodriguez-iturbe97,dodds2000pa,hack57}.

Consider two basins described by $(L_1,W_1)$ 
and $(L_2,W_2)$ within the same system
where $L_i$ is a characteristic longitudinal basin length
and $W_i$ a characteristic width.
The basins
being allometric means that $(W_1/W_2) = (L_1/L_2)^H$ where $H<1$.
Thus, two length ratios are needed to describe the
allometry of basins.  If we consider basins defined by
stream ordering then we have the Horton-like ratios $R_L$ and $R_W=R_L^H$.
Now, when rescaling an entire basin,
streams roughly aligned with a basin's
length will rescale with the factor $R_L$ and those 
perpendicular to the basin's axis will rescale 
differently with $R_W$.
This creates a conundrum: how can basins
be allometric ($R_L \ne R_W$) and yet individual streams 
be self-similar ($R_L = R_W$) as implied by Horton's laws?

We contend the answer is that allometry must be restricted to
\textit{directed networks} and that self-similarity of basins must
hold for \textit{non-directed networks}.  
This is in agreement with Colaiori \etal~\cite{colaiori97}
who also distinguish
between self-similar and allometric river basins
although we stress here the qualification of directedness.
Directed networks have a global direction of flow in which
the direction of each individual stream flow 
has a positive component.  
A basic example is the random model
of Scheidegger~\cite{scheidegger67} which we describe 
below.
For a directed network, $R_L=R_l$, and the rescaling
of basin sizes matches up with the rescaling of
stream lengths regardless of how the basin's width rescales since
all streams are on average aligned with the global
direction of flow.
Hence, our premise that streams rescale in a self-similar way
is general enough to deal with systems whose basins rescale in 
an allometric fashion.

In considering the allometry of basins,
we must also address
the additional possibility
that individual stream lengths
may scale non-trivially with basin length.
In this case, the
main stream length $l$ would vary with the
longitudinal basin length $L$ as $l \propto L^d$.
This is typically a weak dependence with
$1.0 < d < 1.15$~\cite{maritan96a,tarboton90}.
Note that Horton's laws still apply in this case.
The exponent $d$ plays a part in determining whether or not
a basin scales allometrically.
The exponent $H$ introduced in the discussion
of basin allometry can be found in terms of Horton's
ratios (or equivalently Tokunaga's parameters) and $d$
as $H = d\ln{R_n}/\ln{R_l}-1$~\cite{dodds99pa}.

Thus, for a directed network $d=1$
and $H\le1$ (e.g., Scheidegger~\cite{scheidegger67})
whereas for undirected, self-similar networks 
$H=1$ and $d \ge 1$ (e.g., random undirected networks~\cite{manna92,manna96}).
River networks are in practice
often neither fully directed or undirected.
Scaling laws observed in such cases 
will show deviations from pure scaling that
may well be gradual and difficult to detect~\cite{dodds2000ua}.

\section{Tokunaga distributions}
\label{sec:tokunaga.generalization}

The laws of Tokunaga and Horton
relate averages of quantities.
In the remainder of this paper, we 
investigate the underlying distributions from which
these averages are made.  
We are able to find general scaling forms of a number
of distributions and in many cases also identify the basic form
of the relevant scaling function.

\begin{figure}[htb!]
  \begin{center}
    \epsfig{file=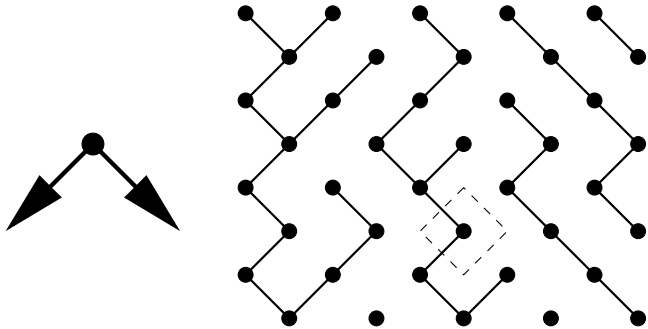,width=.48\textwidth}
    \caption[Scheidegger model of random, directed networks]{
      Scheidegger's random directed networks.
      Sites are arranged on a triangular lattice and
      stream flow is directed down the page.  At each site,
      the flow direction is randomly chosen to be in one
      of the directions shown on the left.  The dashed box indicates
      the area ``drained'' by the local site.
      }
    \label{fig:tokunaga.scheidchoice2}
  \end{center}
\end{figure}

To aid and motivate our investigations, we examine,
as we have done in both~\cite{dodds2000ua} and~\cite{dodds2000ub},
a simple model of directed
random networks that was first
introduced by Scheidegger~\cite{scheidegger67}
Since we make much of use this model in the present
work, we provide a self-contained discussion.
Consider the triangular lattice of sites
oriented as in Figure~\ref{fig:tokunaga.scheidchoice2}.
At each site of the
lattice a stream flow direction 
is randomly chosen between the two
possible diagonal directions shown.
It is therefore trivial to generate
the model on a large scale, allowing
for a thorough investigation of its river network statistics.
The small, tilted box with a dashed boundary represents
the area drained by the enclosed site.
As with many discrete-space models, 
the details of the underlying lattice are unimportant.
On a square lattice, the model's streams would have three
choices of flow, two diagonals and straight down the page.
However, the choice of a triangular lattice does simplify implementation
and calculation of statistics.  
For example, only one tributary can exist at
each site along a stream and stream paths and basin boundaries
are precisely those of the usual 
discrete-space random walk~\cite{feller68I}.

Since random walks are well understood, 
the exponents of many river network scaling laws are exactly
known for the Scheidegger 
model~\cite{takayasu88,takayasu89a,takayasu91,huber91}
and analogies may also be drawn with the
Abelian sandpile model~\cite{dhar99}.
For example, a basin's
boundaries being random walks means that a basin of 
length $L$ will typically have a width $W \propto L^{1/2}$ 
which gives $H=1/2$.
Since the network is directed,
stream length is the same as basin length, $l = L$, so
we trivially have $d=1$.  Basin area $a$ is estimated
by $WL \propto L^{3/2} \propto l^{3/2}$ so $l \propto a^{2/3}$ giving
Hack's law with an exponent of $2/3$~\cite{hack57}.

Nevertheless, the Tokunaga parameters and the
Horton ratios are not known analytically.
Estimates from previous work~\cite{dodds99pa} find
$T_1 \simeq 1.35$, $R_l = R_T  \simeq 3.00$ and $R_n \simeq 5.20$.
Data for the present analysis
was obtained on $L=10^4$ by $W=3\times10^3$
lattices with periodic boundaries.  
Given the self-averaging present in any 
single instance on these networks,
ensembles of 10 were deemed sufficient.

We first examine
the distributions of Tokunaga ratios $\Tmunu$ and observe a strong
link to the underlying distribution of $\okellsnum{\mu}$.
Both are well described by exponential distributions.
To understand this link, we next consider the distances between neighboring
side streams of like order.  This provides a measure of fluctuations
in drainage density and again, exponential distributions
appear.  
We are then in a position to develop theory for the 
joint probability distribution 
between the Tokunaga ratios and stream segment lengths
and, as a result, the
distribution for the quantity $\Tmunu/\okellsnum{\mu}$ and its inverse.
In the limit of large $\mu$, the $\Tmunu$ are effectively
proportional to $\okellsnum{\mu}$ and all fluctuations of the former
exactly follow those of the latter.

All investigations are initially carried out for the Scheidegger model
where we may generate statistics of ever-improving
quality.  We find the same forms for all distributions
for the Mississippi data (and for
other river networks not presented here) and provide
some pertinent examples.  
Perhaps the most significant
benefit of the simple Scheidegger model is its ability to
provide clean distributions whose form we can then search
for in real data.

\begin{figure}[tbh!]
  \begin{center}
    \epsfig{file=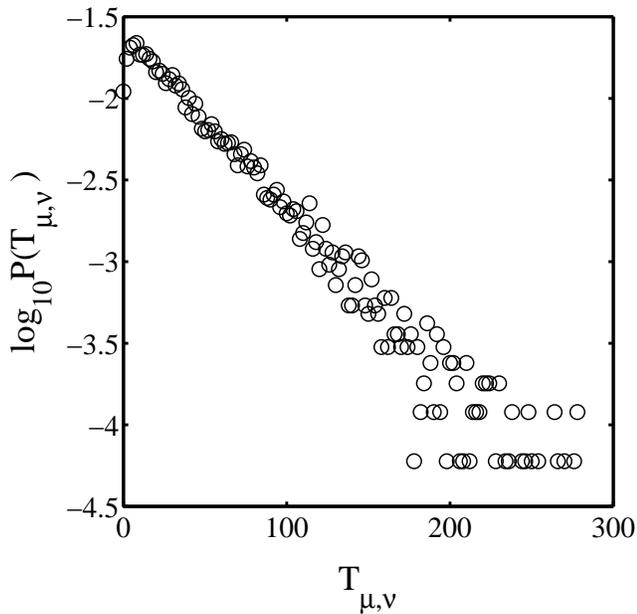,width=0.48\textwidth}
    \caption[Generalized Tokunaga distribution for the Scheidegger model]{
      An example of a generalized Tokunaga distribution
      for the Scheidegger model.
      The Tokunaga ratio $\Tmunu$ is the number of side streams
      of order $\nu$ entering an absorbing stream of order $\mu$.
      For this particular example $\mu=6$ and $\nu=2$.
      The form is exponential and is a result of variations
      in stream segment length rather than significant fluctuations
      in side stream density.
      }
    \label{fig:tokunaga.sche_Tok}
  \end{center}
\end{figure}

Figure~\ref{fig:tokunaga.sche_Tok}
shows the distribution of the number
of order $\nu=2$ side streams entering
an order $\mu=6$ absorbing stream
for the Scheidegger model.
At first, it may seem surprising
that this is not a single-peaked distribution
centered around $\tavg{\Tmunu}$
dying off for small and
large values of $\Tmunu$.

\begin{figure}[tb!]
  \begin{center}
    \ifthenelse{\boolean{@twocolumn}}
    {
    \begin{tabular}{c}
      \textbf{(a)} \\ 
      \epsfig{file=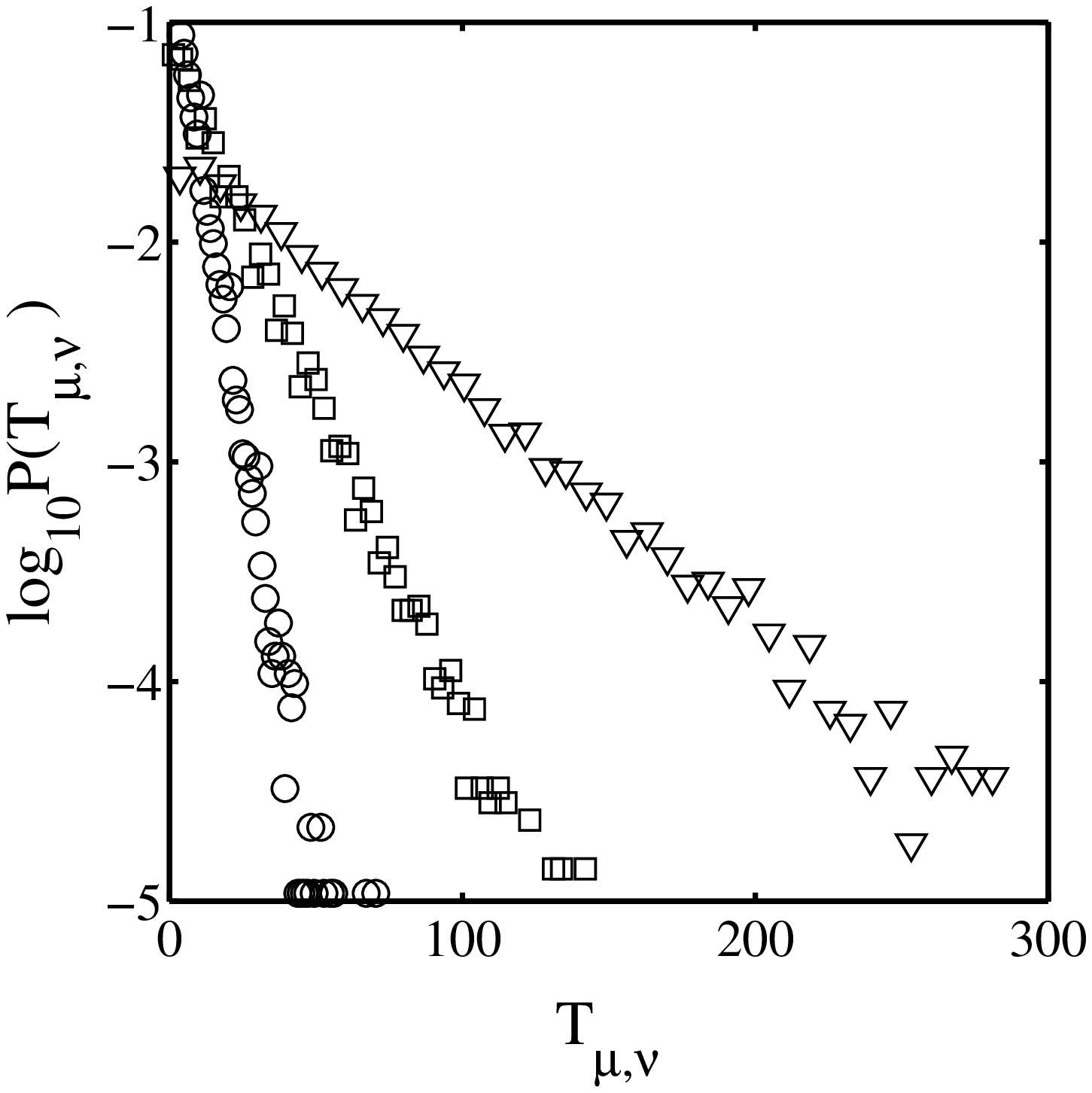,width=0.48\textwidth} \\
      \textbf{(b)} \\
      \epsfig{file=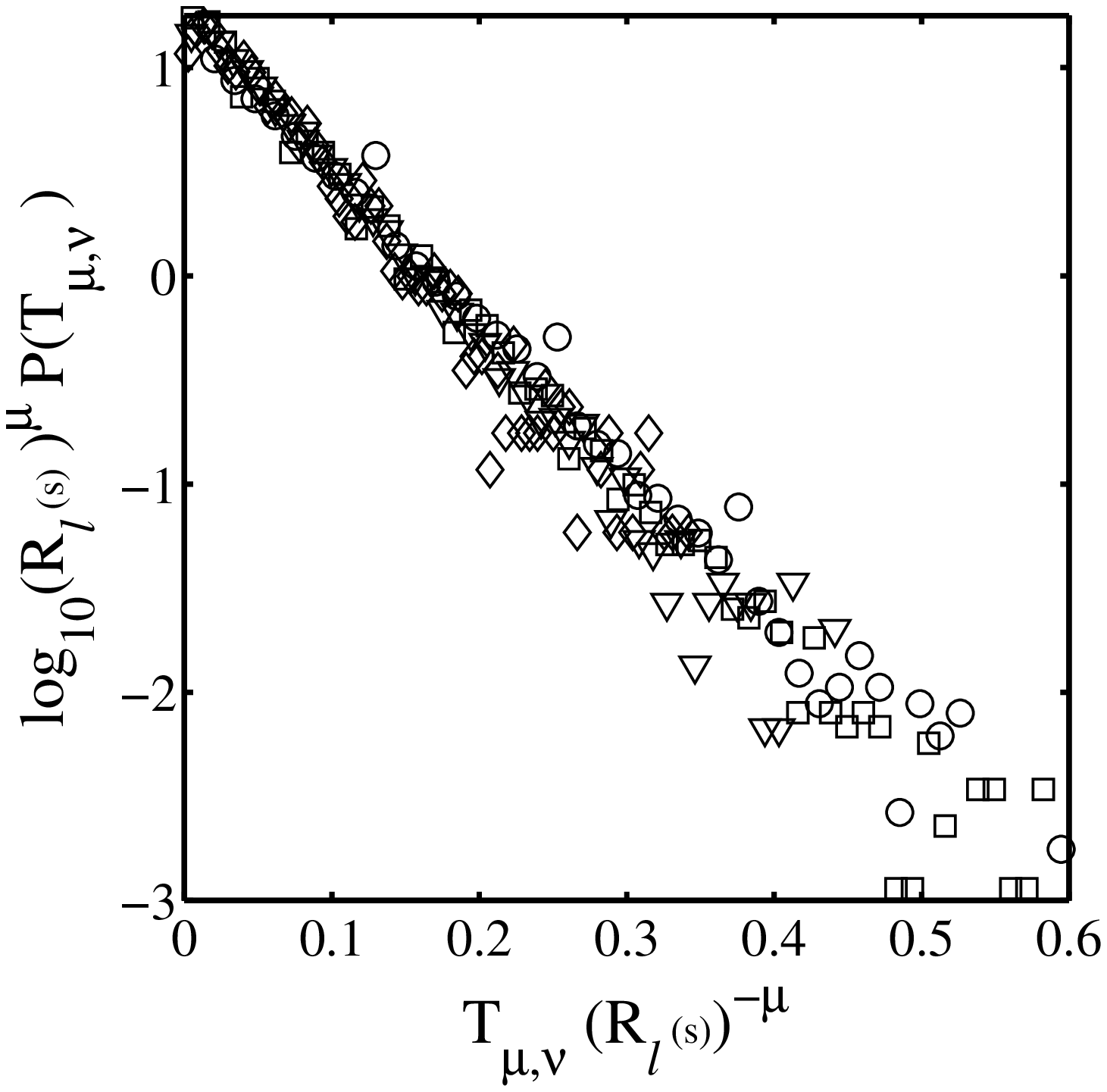,width=0.48\textwidth}
    \end{tabular}
      }
    {
    \begin{tabular}{cc}
      \textbf{(a)} & \textbf{(b)} \\
      \epsfig{file=figsche_Tok_omega_uncollapse_noname.ps,width=0.48\textwidth} &
      \epsfig{file=figsche_Tok_omega_collapse_noname.ps,width=0.48\textwidth}
    \end{tabular}
    }
    \caption[Tokunaga distributions for varying absorbing stream order $\mu$ for the Scheidegger network]{
      Distributions for Tokunaga ratios for varying
      orders of absorbing stream and fixed side stream order of $\nu=2$
      for the Scheidegger network.
      In (a), examples of $\Tmunu$ distributions for
      absorbing stream order $\mu=4$ (circles), $\mu=5$ (squares)
      and $\mu=6$ (triangles).  
      In (b), these distributions,
      as well as the $\mu=7$ 
      case, are rescaled according
      to equation~\req{eq:tokunaga.tokgen1}.
      The resulting ``data collapse''
      gives a single distribution.
      For the Scheidegger model, $R_\okell \simeq 3.00$.
      }
    \label{fig:tokunaga.sche_Tok_omega_collapse}
  \end{center}
\end{figure}

The distribution of $\Tmunu$ 
in Figure~\ref{fig:tokunaga.sche_Tok}
is clearly well described
by an exponential distribution.
This can also be seen upon inspection of
Figures~\ref{fig:tokunaga.sche_Tok_omega_collapse}(a)
and~\ref{fig:tokunaga.sche_Tok_omega_collapse}(b).
Figure~\ref{fig:tokunaga.sche_Tok_omega_collapse}(a)
shows normalized distributions of $\Tmunu$ for
$\nu=2$ and varying absorbing stream order $\mu=4$, 5 and 6.
These distributions (plus the one for absorbing
stream order $\mu=7$) are rescaled and
presented in Figure~\ref{fig:tokunaga.sche_Tok_omega_collapse}(b).
The single form thus obtained suggests
a scaling form of the $\Tmunu$ distribution is given by
\begin{equation}
  \label{eq:tokunaga.tokgen1}
  P(\Tmunu) = (R_\okell)^{-\mu}
  F\left[ \Tmunu (R_\okell)^{-\mu}  \right].
\end{equation}
where $F$ is an exponential scaling function.
However, this only accounts for variations in $\mu$,
the order of the absorbing stream.

\begin{figure}[tb!]
  \begin{center}
    \ifthenelse{\boolean{@twocolumn}}
    {
    \begin{tabular}{c}
      \textbf{(a)} \\
      \epsfig{file=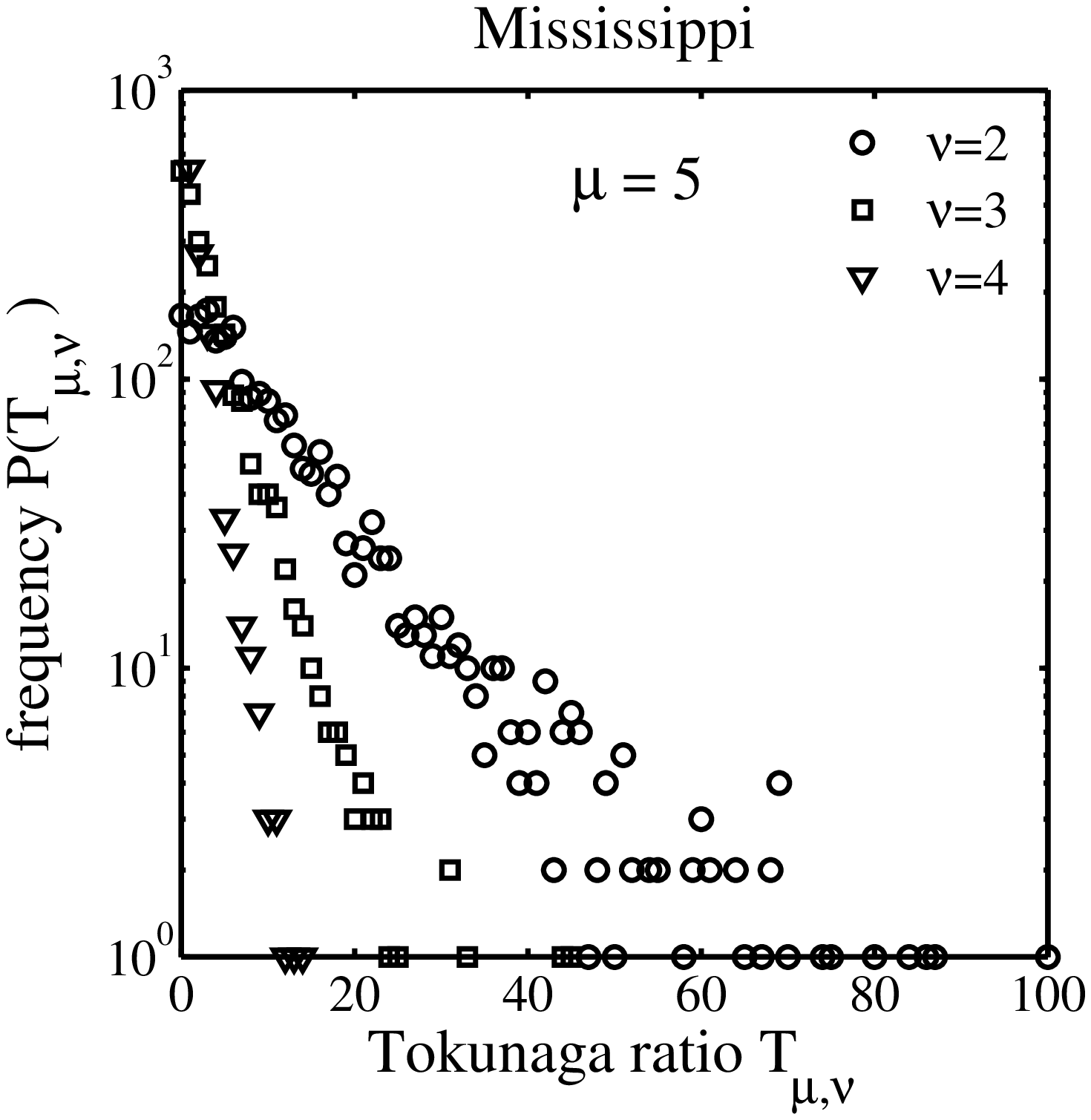,width=0.48\textwidth} \\
      \textbf{(b)} \\
      \epsfig{file=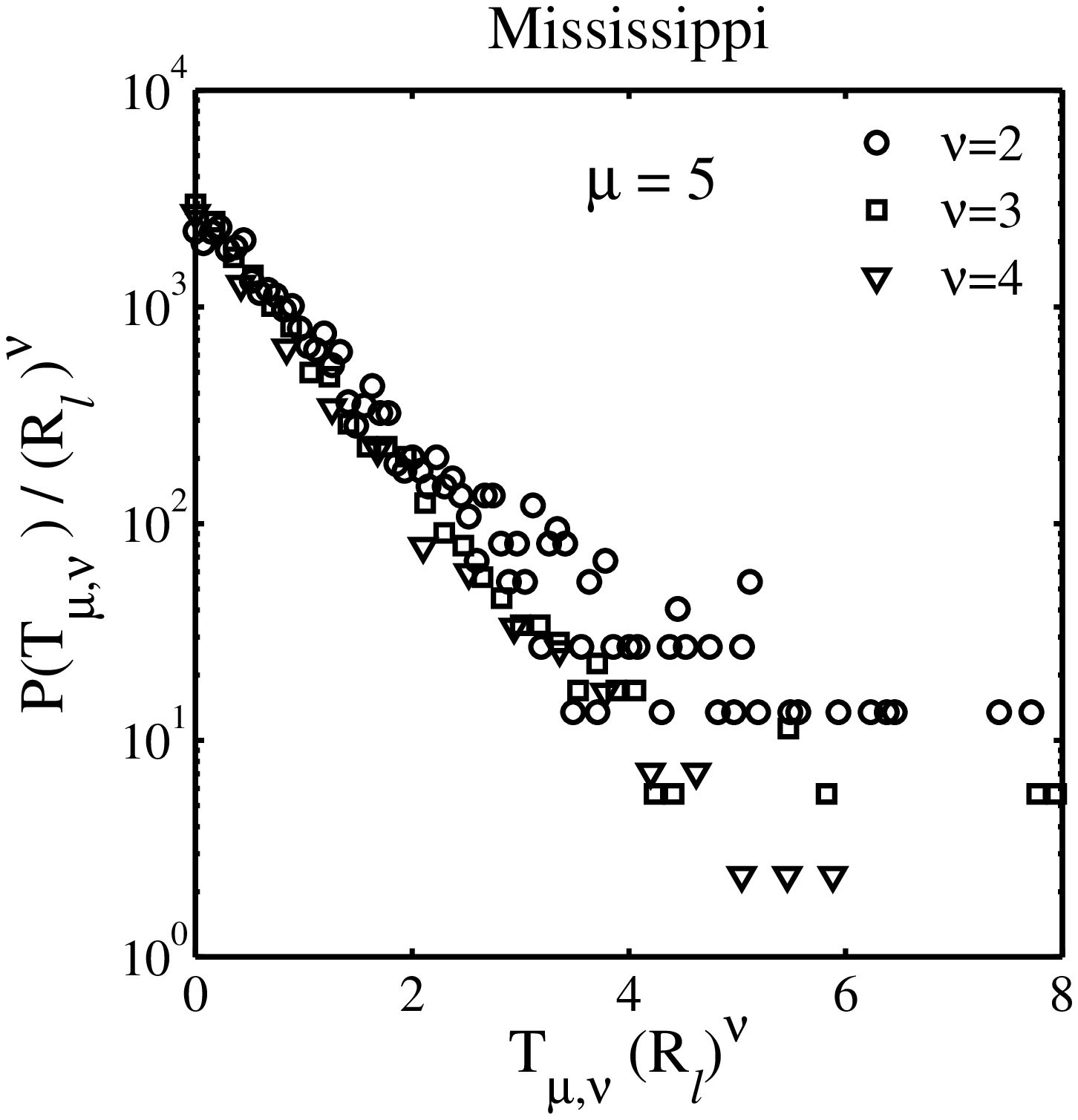,width=0.48\textwidth} 
    \end{tabular}
    }
    {
    \begin{tabular}{cc}
      \textbf{(a)} & \textbf{(b)} \\
      \epsfig{file=figtokdist_order1_b_noname.ps,width=0.48\textwidth} &
      \epsfig{file=figtokdist_order1_collapse_b_noname.ps,width=0.48\textwidth}
    \end{tabular}
    }
    \caption[Tokunaga distributions for the Mississippi for varying side stream order $\nu$]{
      Tokunaga distributions for varying
      side stream orders for the Mississippi river basin.
      In both (a) and (b), the absorbing stream order is $\mu=5$ and the
      side stream orders are $\nu=2$ (circles),
      $\nu=3$ (squares) and $\nu=4$ (triangles).
      The raw distributions are shown in (a).
      In (b) the distributions are rescaled
      as per equation~\req{eq:tokunaga.tokgen2}.
      For the Mississippi, the ratio 
      is estimated to be 
      $R_\okell \simeq 2.40$~\protect\cite{dodds2000ub}.
      }
    \label{fig:tokunaga.tokdist_order1_b}
  \end{center}
\end{figure}

Figures~\ref{fig:tokunaga.tokdist_order1_b}(a)
and ~\ref{fig:tokunaga.tokdist_order1_b}(b)
show that a similar rescaling 
of the distributions may
be effected when $\nu$ is varied.  
In this case, the
data is for the Mississippi.
The rescaling is now by $R_\okell$
rather than $R_\okell^{-1}$ and equation~\req{eq:tokunaga.tokgen1}
is improved to give
\begin{equation}
  \label{eq:tokunaga.tokgen2}
  P(\Tmunu) = (R_\okell)^{\mu-\nu-1}
  P_T\left[ \Tmunu / (R_\okell)^{\mu-\nu-1}  \right].
\end{equation}
The function $P_T$ is a normalized exponential distribution
independent of $\mu$ and $\nu$,
\begin{equation}
  \label{eq:tokunaga.PT}
  P_T(z) = \frac{1}{\xi_t} e^{-z/\xi_t},
\end{equation}
where $\xi_t$ is the characteristic number
of side streams of one order lower than
the absorbing stream, i.e., $\xi_t = \tavg{T_1}$.
For the Mississippi, we observe
$\xi_t \simeq 1.1$
whereas for the Scheidegger model,
$\xi_t \simeq 1.35$.
As expected, the Tokunaga distribution is dependent only on
$k=\mu-\nu$ so we can write
\begin{equation}
  \label{eq:tokunaga.tokgen3}
  P(T_k) = (R_\okell)^{k-1}
  P_T\left[ \Tmunu / (R_\okell)^{k-1} \right].
\end{equation}
with $P_T$ as above.

\section{Distributions of stream segment lengths and randomness}
\label{sec:tokunaga.streamsegs}

\begin{figure}[tb!]
  \begin{center}
    \ifthenelse{\boolean{@twocolumn}}
    {
    \begin{tabular}{c}
      \textbf{(a)} \\
      \epsfig{file=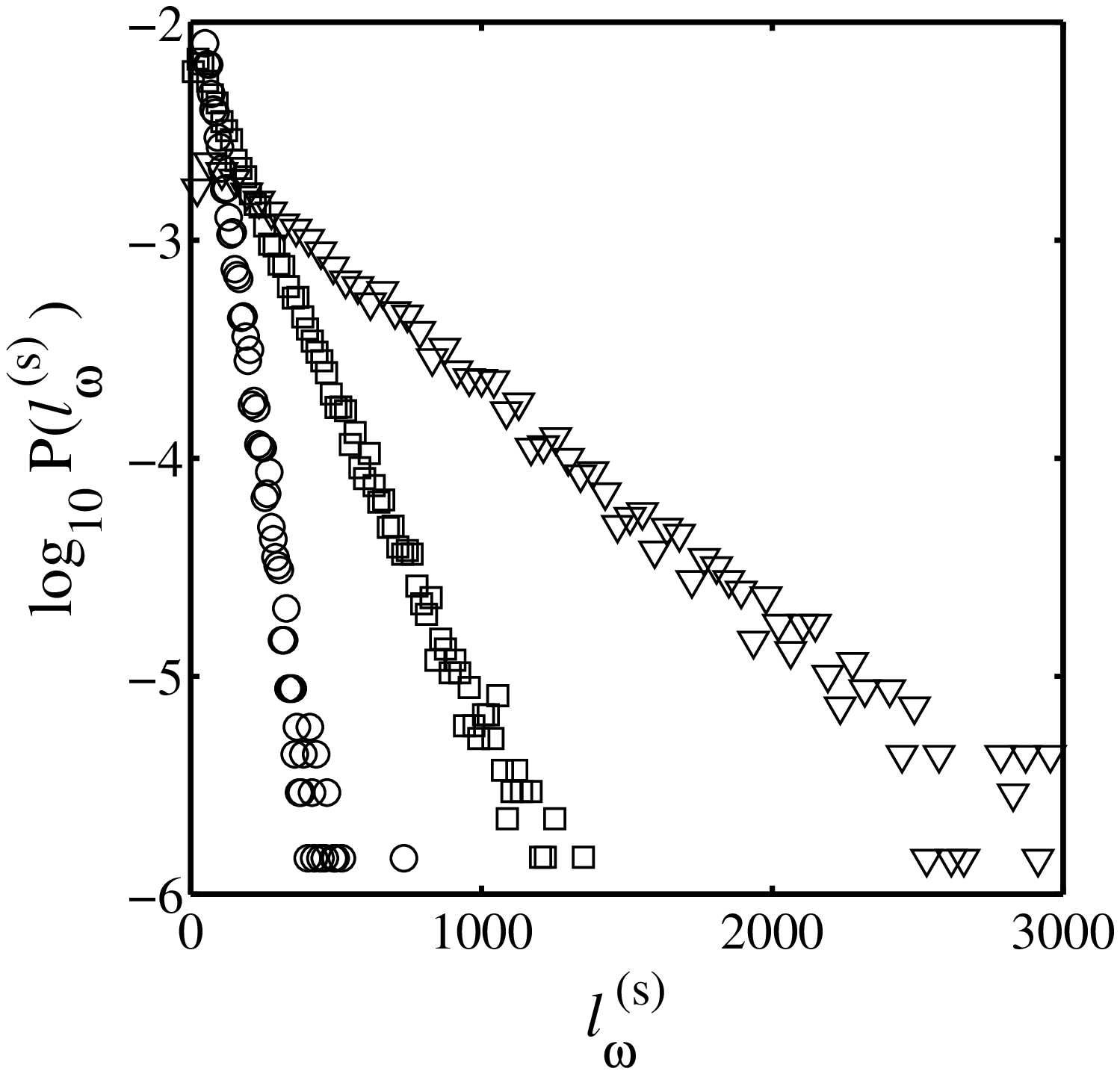,width=0.48\textwidth} \\
      \textbf{(b)} \\
      \epsfig{file=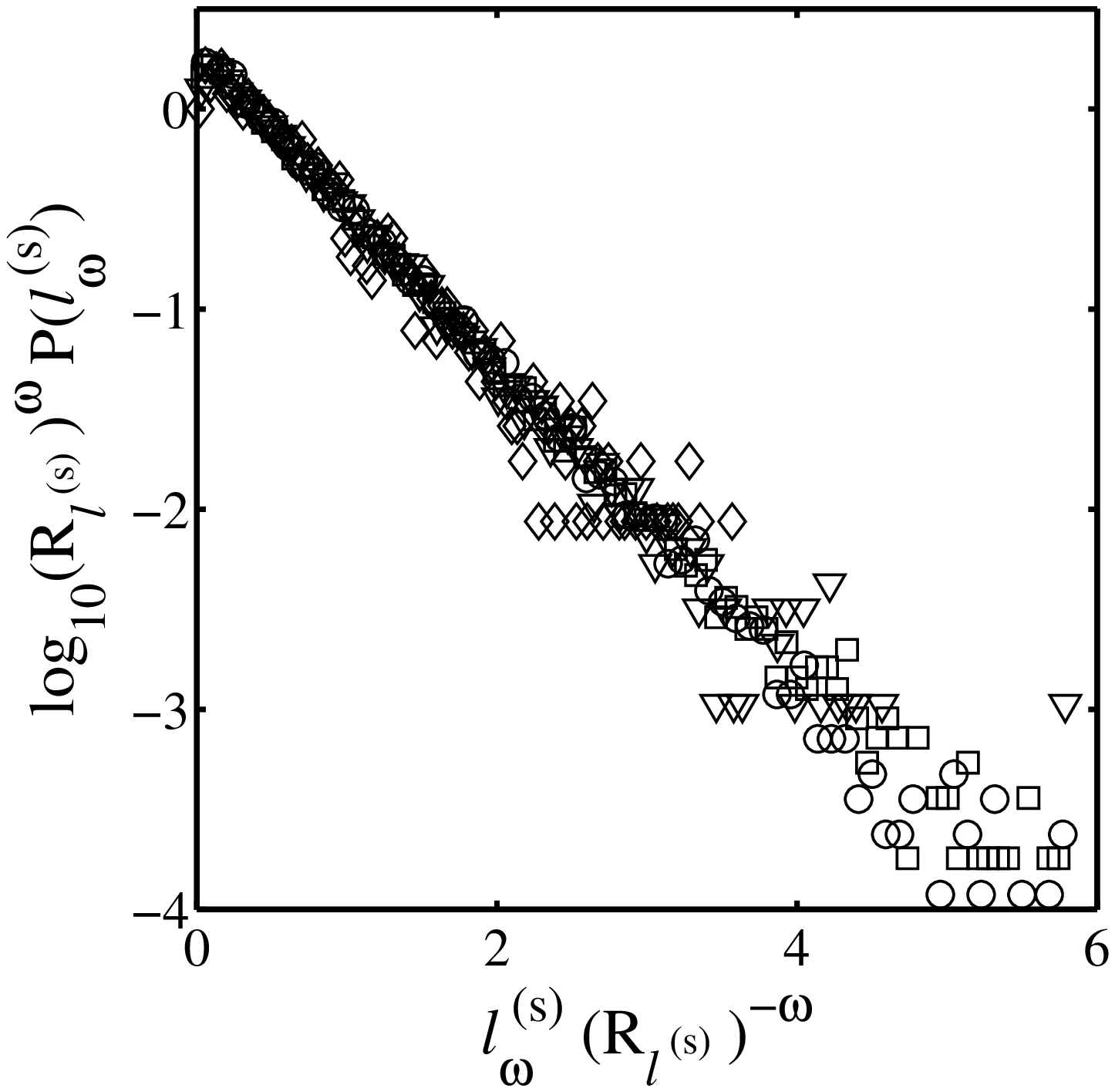,width=0.48\textwidth}
    \end{tabular}
    }
    {
    \begin{tabular}{cc}
      \textbf{(a)} & \textbf{(b)} \\
      \epsfig{file=figsche_ellomega_uncollapse_noname.ps,width=0.48\textwidth} &
      \epsfig{file=figsche_ellomega_collapse_noname.ps,width=0.48\textwidth}
    \end{tabular}
    }
    \caption[Stream segment length distributions for the Scheidegger model]{
      Stream segment length distributions for varying stream order
      for the Scheidegger model.  
      Lengths are in units the lattice spacing.
      Shown in (a) are raw distributions for $\omega=4$ (circles),
      $\omega=5$ (squares) and $\omega=6$.
      The linear forms on the semilogarithmic axes indication
      these distributions are well approximated by 
      exponentials~\cite{dodds2000ub}.
      In (b), the distributions in (a) plus the 
      distribution for $\okellsnum{7}$ (diamonds) 
      are rescaled using equation~\req{eq:tokunaga.ellgen}.
      }
    \label{fig:tokunaga.sche_ellomega_collapse}
  \end{center}
\end{figure}

As we have suggested,
the distributions of the Tokunaga ratios depend strongly
on the distributions of stream segment lengths.
Figure~\ref{fig:tokunaga.sche_ellomega_collapse}
is the indicates why this is so.
The form of the underlying distribution 
is itself exponential.  We have already examined this fact
extensively in~\cite{dodds2000ub} and here we
develop its relationship with the Tokunaga distributions.

Figures~\ref{fig:tokunaga.sche_ellomega_collapse}(a)
and~\ref{fig:tokunaga.sche_ellomega_collapse}(b)
show that the distributions of $\okellsnum{\mu}$ 
can be rescaled in the same way as the Tokunaga
distributions.
Thus, we write the distribution for stream segment
lengths as~\cite{dodds2000ub}
\begin{equation}
  \label{eq:tokunaga.ellgen}
  P(\okellsnum{\mu}) = (R_\okell)^{-\mu+1}
  P_\okell\left[ \okellsnum{\mu} / (R_\okell)^{-\mu+1}  \right].
\end{equation}
As for $P_T$, the function $P_\okell$ is a normalized exponential distribution
\begin{equation}
  \label{eq:tokunaga.Pell}
  P_\okell(z) = \frac{1}{\xi_\okell} e^{-z/\xi_\okell},
\end{equation}
where, in a strictly self-similar network,
$\xi_\okell$ is the characteristic 
length of first-order stream segments,
i.e., $\xi_\okell = \tavg{\okellsnum{1}}$.
(Note that in~\cite{dodds2000ub} we use $\xi$
for $\xi_\okell$ for ease of notation).
We qualify this by requiring the network to be
exactly self-similar because in most models
and all real networks this is certainly not
the case.  As should be expected,
there are deviations from scaling
for the largest and smallest orders.
Therefore, $\xi_\okell$ is 
the characteristic size of a first-order stream
as determined by scaling down the average lengths of
those higher order streams
that are in the self-similar 
structure of the network.
It is thus in general
different from $\tavg{\okellsnum{1}}$.

We therefore see that the distributions of $\Tmunu$
and $\okellsnum{\mu}$ are both exponential in form.
Variations in $\okellsnum{\mu}$ largely govern
the possible values of the $\Tmunu$.
However, $\Tmunu$ is still only 
proportional to $\okellsnum{\mu}$ on average
and later on we will explore the joint distribution
from which these individual exponentials arise.

The connection between the characteristic number $\xi_t$
and the length-scale $\xi_\okell$ follows from 
equations~\req{eq:tokunaga.hortslaws_ell},
\req{eq:tokunaga.tokellconnect},
and~\req{eq:tokunaga.hortondd2}:
\begin{equation}
  \label{eq:tokunaga.xiconnect}
  \xi_t = \rho_1 R_\okell \xi_\okell.
\end{equation}
This presumes exact scaling of drainage densities
and in the case where this is not so, $\rho_1$
would be chosen so that $(R_\okell)^{\nu-1} \rho_1$
most closely approximates the higher order $\rho_\nu$.

We come to an important interpretation of the exponential
distribution as a composition of independent probabilities.
Consider the example of stream segment lengths.
We write $\tilde{p}_{\mu}$ as the probability that a stream segment 
of order $\mu$ meets with
(and thereby terminates at)
a stream of order at least $\mu$.
For simplicity, we assume only
one side stream or none may join a stream at any site.
We also take the lattice spacing $\alpha$
to be unity so that stream lengths are integers
and therefore equate with the number of links between
sites along a stream.
For $\alpha \neq 1$,
derivations similar to below will apply
with $\okellsnum{\mu}$ replaced by $[\okellsnum{\mu}/\alpha]$,
where $[\cdot]$ denotes rounding to the nearest integer.
Note that extra complications arise when 
the distances between neighboring sites are not uniform.

Consider a single instance of an order $\mu$ stream segment.
The probability of this segment having a
length $\okellsnum{\mu}$ is given by
\begin{equation}
  \label{eq:tokunaga.probell}
  P(\okellsnum{\mu}) = \tilde{p}_{\mu} 
  ( 1 - \tilde{p}_{\mu})^{\okellsnum{\mu}}.
\end{equation}
where $\tilde{p}_{\mu}$ is the probability that
an order an order $\mu$ stream segment terminates
on meeting a stream of equal or higher order.
We can re-express the above equation as
\begin{equation}
  \label{eq:tokunaga.probell2}
  P(\okellsnum{\mu}) \simeq \tilde{p}_{\mu} 
  \exp\{ -\okellsnum{\mu}
  \ln(1-\tilde{p}_{\mu})^{-1} \},
\end{equation}
and upon inspection of equations~\req{eq:tokunaga.ellgen}
and~\req{eq:tokunaga.Pell} we make the identification
\begin{equation}
  \label{eq:tokunaga.xipconnect}
  (R_\okell)^{\mu-1} \xi_\okell = [-\ln(1-\tilde{p}_{\mu})]^{-1},
\end{equation}
which has the inversion
\begin{equation}
  \label{eq:tokunaga.xipconnect2}
  \tilde{p}_{\mu} = 1 - e^{-1/(R_\okell)^{\mu-1}\xi_\okell}.
\end{equation}
For $\mu$ sufficiently large
such that $\tilde{p}_{\mu} \ll 1$, we have the simplification
\begin{equation}
  \label{eq:tokunaga.xipconnect3}
  \tilde{p}_{\mu} \simeq 1/(R_\okell)^{\mu-1}\xi_\okell.
\end{equation}
We see that the probabilities satisfy the Horton-like
scaling law
\begin{equation}
  \label{eq:tokunaga.pscaling}
  \tilde{p}_{\mu}/\tilde{p}_{\mu-1} = 1/R_\okell.  
\end{equation}
Thus, we begin to see the element of
randomness in our expanded description of
network architecture.  
The termination
of a stream segment by meeting a larger branch
is effectively a spatially random process.

\section{Generalized drainage density}
\label{sec:tokunaga.drainagedensity}

\begin{figure}[tb!]
  \begin{center}
    \epsfig{file=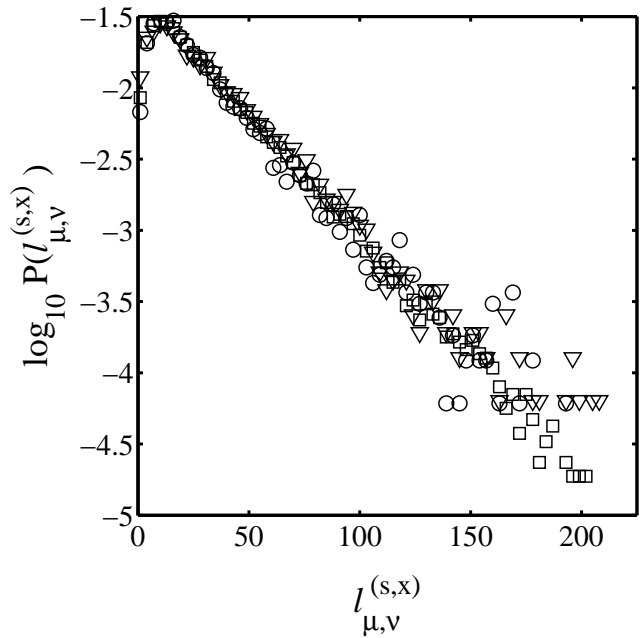,width=0.48\textwidth}
    \caption[Inter-tributary length distributions for the Scheidegger model]{
      A comparison of inter-tributary length distributions
      for the Scheidegger model.  
      The example here is for order $\mu=6$ absorbing streams
      and order $\nu=3$ side streams.
      Note that no rescaling of the distributions has been performed.
      The three length variables here 
      correspond to x=b, x=i and x=e, i.e., $\okellb$ (circles)
      $\okelli$ (squares), and $\okelle$ (triangles).
      These are
      the beginning, internal and end distances
      between entering side branches, defined fully in the text.
      No quantitative difference between these three lengths
      is observed.
      }
    \label{fig:tokunaga.sche_ellibe_comparison}
  \end{center}
\end{figure}

Having observed the similarity of the distributions
of $\Tmunu$ and $\okellsnum{\mu}$,
we proceed to examine the exact nature
of the relationship beteween the two.
To do so, we introduce three new measures of stream length.
These are $\okellb$, 
the distance from the beginning of
an order $\mu$ absorbing stream to the first order $\nu$ 
side stream; 
$\okelli$, the distance between any two adjacent
internal order $\nu$ side streams along an order
$\mu$ absorbing stream; 
and $\okelle$, the distance from
the last order $\nu$ side stream to the end
of an order $\mu$ absorbing stream.
By analysis of these inter-tributary lengths, we
will be able to discern the distribution of side stream
location along absorbing streams.  
This leads directly
to a more general picture of drainage density
which we fully expand upon in the following section.

\begin{figure}[tbp!]
  \begin{center}
    \ifthenelse{\boolean{@twocolumn}}
    {
    \begin{tabular}{c}
      \textbf{(a)} \\
      \epsfig{file=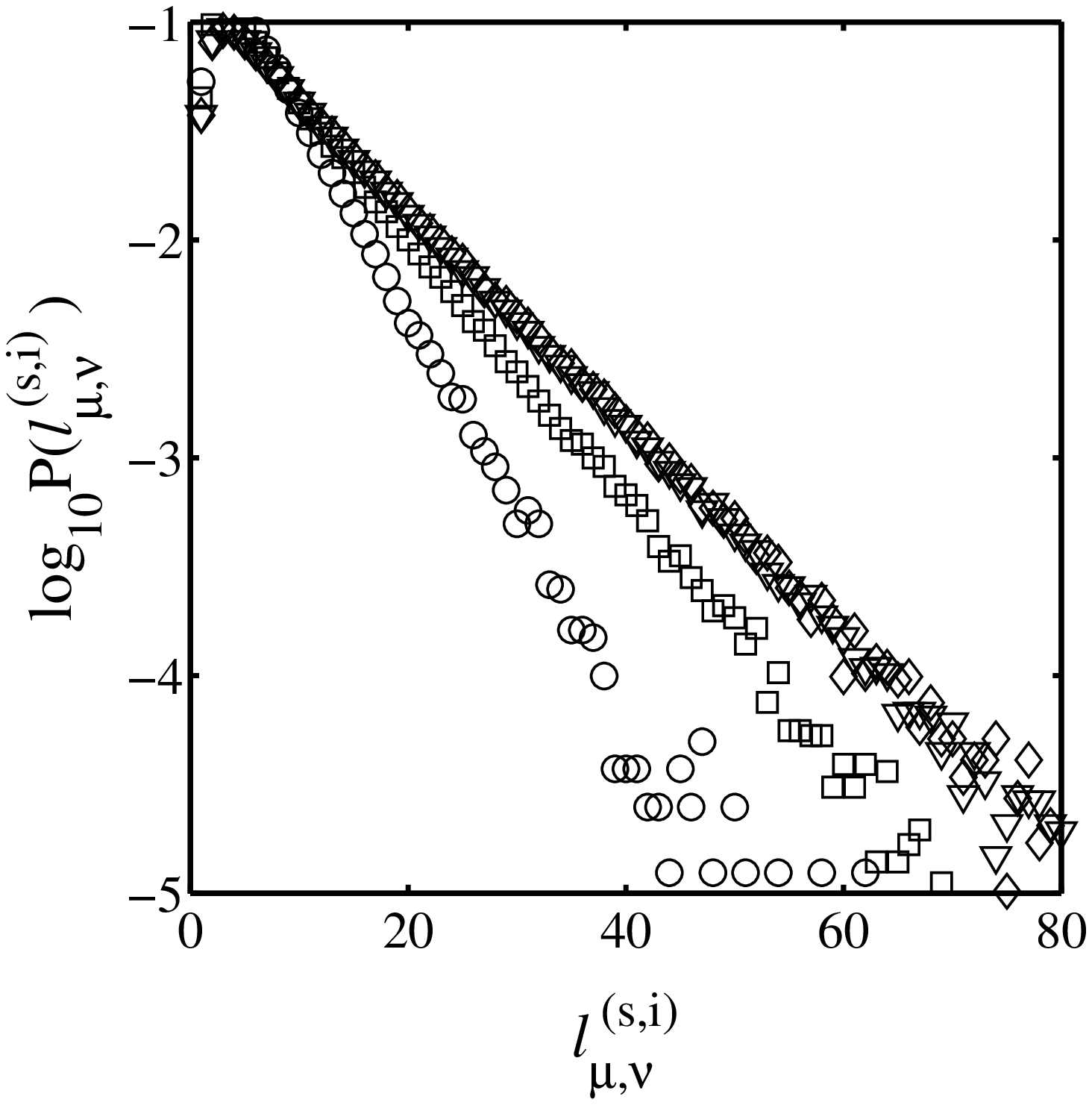,width=0.48\textwidth} \\
      \textbf{(b)} \\
      \epsfig{file=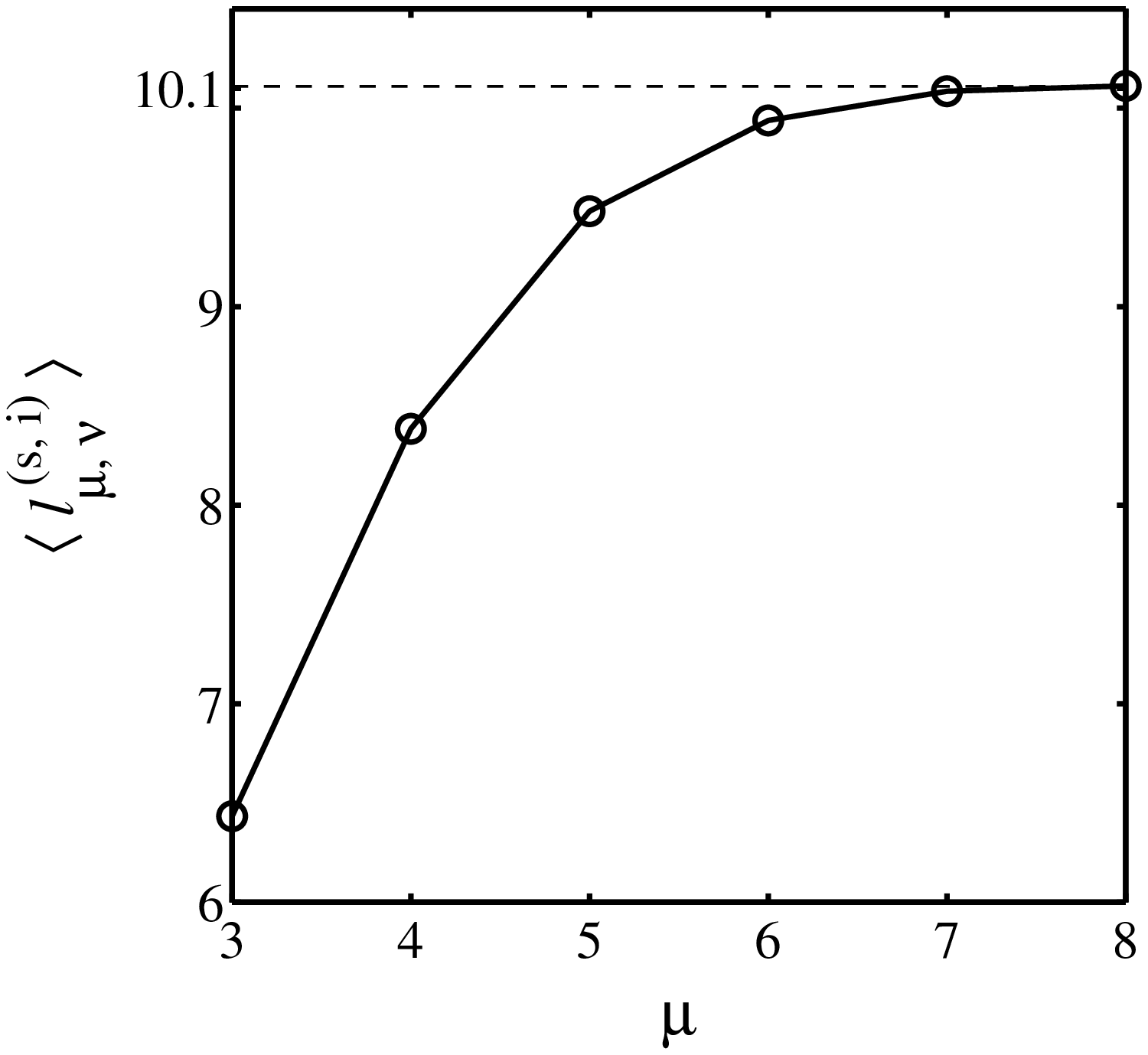,width=0.48\textwidth}
    \end{tabular}
    }
    {
    \begin{tabular}{cc}
      \textbf{(a)} & \textbf{(b)} \\
      \epsfig{file=figsche_elli_omegavary_noname.ps,width=0.48\textwidth} &
      \epsfig{file=figsche_meanelli_noname.ps,width=0.48\textwidth}
    \end{tabular}
    }
    \caption[Asymptotic form of distributions of inter-tributary lengths for the Scheidegger model]{
      An examination of 
      the asymptotic behavior of distributions 
      of internal inter-tributary lengths.
      The data here is for the Scheidegger model
      for the case of fixed side stream order $\nu=2$.
      The plot in (a) shows
      distributions for absorbing stream order
      $\mu=3$ (circles), $\mu=4$ (squares),
      $\mu=6$ (triangles), and $\mu=8$ (diamonds).
      As $\mu$ increases, the distributions, which
      are all individually exponential, tend towards
      a fixed exponential distribution.
      Since lower order
      stream segments have typically smaller lengths, 
      they statistically block larger values of
      $\okelli$, reducing the extent of the 
      distribution tails for low $\mu$.
      This is further evidenced in (b) which provides
      a plot of $\tavg{\okelli}$, the mean inter-tributary
      stream length, as a function of $\mu$ with $\nu=2$.
      These mean values approach
      $\tavg{\okellinum{2}} = 1/\rho_2 \simeq 10.1$ where $\rho_2$ is the density
      of second-order side streams.
      }
    \label{fig:tokunaga.sche_elli_omegavary}
  \end{center}
\end{figure}

Figure~\ref{fig:tokunaga.sche_ellibe_comparison}
compares normalized distributions of $\okellb$, $\okelli$ and $\okelle$
for the Scheidegger model.  The data is for the distance
between order $\nu=3$ side streams entering order $\mu=6$
absorbing streams.
Once again, the distributions
are well approximated by exponential distributions.
Moreover, they are indistinguishable.
This indicates, at least for the Scheidegger model,
that drainage density is independent 
of relative position of tributaries along an absorbing
stream.  

We now consider the effect on the distribution 
of internal inter-tributary distances $\okelli$
following from variations in $\mu$, the order of the
absorbing stream.  
Figure~\ref{fig:tokunaga.sche_elli_omegavary}(a)
provides a comparison of $\okelli$ distributions for $\nu=2$ and
$\mu=3$ through $\mu=8$.  As $\mu$ increases, the distributions
tend towards a limiting function.  With increasing $\mu$
we are, on average, sampling absorbing streams of greater
length and the full range of $\okelli$ becomes
accordingly more accessible.  
This approach to a fixed distribution
is reflected in the means of the distributions
in Figure~\ref{fig:tokunaga.sche_elli_omegavary}(a).
Shown in Figure~\ref{fig:tokunaga.sche_elli_omegavary}(b),
the means $\tavg{\okelli}$ for $\nu=2$ approach
a value of around $10.1$.  The corresponding
density of second-order streams for
the Scheidegger model is thus
$\rho_2 = 1/\tavg{\okellinum{2}} \simeq 0.01$.
Higher drainage densities follow from
equation~\req{eq:tokunaga.hortondd2}.
However, since
deviations occur for small $\nu$, there will
also be an approach to uniform scaling to consider
with drainage density.

\section{Joint variation of Tokunaga ratios and stream segment length}
\label{sec:tokunaga.Tell}

We have so far observed that the individual distributions
of the $\okellsnum{\mu}$ and $\Tmunu$ are exponential
and that they are related via
the side-stream density $\rho_nu$.
However, this is not an exact relationship.
For example,
given a collection of stream segments with a fixed length $\okellsnum{\mu}$
we expect to find fluctuations in the 
corresponding Tokunaga ratios $\Tmunu$.

To investigate this further we now
consider the joint variation of $\Tmunu$ with $\okellsnum{\mu}$
from a number of perspectives.
After discussing the full joint probability distribution
$P(\Tmunu,\okellsnum{\mu})$
we then focus
on the quotient
$v=\Tmunu/\okellsnum{\mu}$ and its reciprocal
$w=\okellsnum{\mu}/\Tmunu$.  
The latter two quantities are measures
of drainage density and 
inter-tributary length for an individual absorbing stream.

\subsection{The joint probability distribution}
We build the joint distribution 
of $P(\Tmunu,\okellsnum{\mu})$ from 
our conception that stream segments
are randomly distributed throughout a basin.
In equation~\req{eq:tokunaga.probell},
we have the probability of 
a stream segment terminating 
after $\okellsnum{\mu}$ steps.
We need to incorporate into this form
the probability
that the stream segment also has
$\Tmunu$ order $\nu$ side streams.
Since we assume placement of these side streams to be random, 
we modify equation~\req{eq:tokunaga.probell} to find
\begin{equation}
  \label{eq:tokunaga.probellT}
  P(\okellsnum{\mu},\Tmunu) = \tilde{p}_{\mu}
  \binom{\okellsnum{\mu} - 1}{\Tmunu}
  p_{\nu}^{\Tmunu}
  ( 1 - p_\nu - \tilde{p}_{\mu})^{\okellsnum{\mu} - \Tmunu - 1},
\end{equation}
where $\tbinom{n}{k} = n!/k!(n-k)!$ is the binomial
coefficient and $p_\nu$ is the probability
of absorbing an order $\nu$ side stream.
The extra $p_\nu$
appears in the last factor $( 1 - p_\nu - \tilde{p}_{\mu})$
because this term is the probability that at a particular site
the stream segment neither
terminates nor absorbs an order $\nu$ side stream.
Also, it is simple to verify that the sum over $\okellsnum{\mu}$
and $\Tmunu$ of the probability 
in equation~\req{eq:tokunaga.probell2} returns unity.

While equation~\req{eq:tokunaga.probell2}
does precisely describe the joint distribution
$P(\okellsnum{\mu},\Tmunu)$, it is somewhat
cumbersome to work with.  We therefore
find an analogous form defined for
continuous rather than discrete variables.
We simplify our notation by writing
$p = p_\nu$, $q = ( 1 - p_\nu - \tilde{p}_{\mu})$
and $\tilde{p} = \tilde{p}_{\mu}$.
We also replace $(\okellsnum{\mu}, \Tmunu)$ by
$(x,y)$ where now $x, y \in \mathbb{R}$.
Note that $0 \le y \le x-1$ since the number
of side streams cannot be greater than
the number of sites within a stream segment.

Equation~\req{eq:tokunaga.probell2} becomes
\begin{equation}
  \label{eq:tokunaga.probxy}
  P(x,y) = N\tilde{p} \,
  \frac{\Gamma(x)}{\Gamma(y+1)\Gamma(x-y)}
  (p)^{y} (q)^{x - y - 1},
\end{equation}
where we have used $\Gamma(z+1) = z!$
to generalize the binomial coefficient.
We have included the normalization $N$
to account for the fact that we have moved
to continuous variables and the resulting
probability may not be cleanly normalized.
Also we must allow that $N=N(p,\tilde{p})$
and we will be able to identify this form more fully later on.
Using Stirling's approximation~\cite{bender78},
that $\Gamma(z+1) \sim \sqrt{2\pi} z^{z+1/2} e^{-z}$,
we then have
\ifthenelse{\boolean{@twocolumn}}
{\begin{eqnarray}
  \label{eq:tokunaga.probxy2}
  P(x,y) & = & N\tilde{p}
  p^{y} q^{x - y - 1}
  \frac{1}{\sqrt{2\pi}}
  \frac{(x-1)^{x-3/2}}{y^{y+1/2} (x-y-1)^{x-y-1/2}} \nonumber \\
  & = & N\frac{\tilde{p}}{\sqrt{2\pi}q}
  p^{y} q^{x - y}
  (x-1)^{-1/2} \nonumber \\
  &   & \times
  \rbfrac{y}{x-1}^{-y-1/2}
  \left(1 - \frac{y}{x-1}\right)^{-x+y+1/2} \nonumber \\
  & \simeq & 
  N' x^{-1/2}
  \left[ F(y/x) \right]^{x}
\end{eqnarray}}
{\begin{eqnarray}
  \label{eq:tokunaga.probxy2}
  P(x,y) & = & N\tilde{p}
  p^{y} q^{x - y - 1}
  \frac{1}{\sqrt{2\pi}}
  \frac{(x-1)^{x-3/2}}{y^{y+1/2} (x-y-1)^{x-y-1/2}} \nonumber \\
  & = & N\frac{\tilde{p}}{\sqrt{2\pi}q}
  p^{y} q^{x - y}
  (x-1)^{-1/2}
  \rbfrac{y}{x-1}^{-y-1/2}
  \left(1 - \frac{y}{x-1}\right)^{-x+y+1/2} \nonumber \\
  & \simeq & 
  N' x^{-1/2}
  \left[ F(y/x) \right]^{x}
\end{eqnarray}}
where we have absorbed $N$ and all terms
involving only $p$ and $\tilde{p}$ into
the prefactor $N'=N'(p,\tilde{p}) = N\tilde{p}/(\sqrt{2\pi}q)$.
We have also assumed $x$ is large
such that $x-1 \simeq x$ and $1 \gg 1/x \simeq 0$.

\begin{figure}[tbp!]
  \begin{center}
    \ifthenelse{\boolean{@twocolumn}}
    {
    \begin{tabular}{c}
      \textbf{(a)} \\
      \epsfig{file=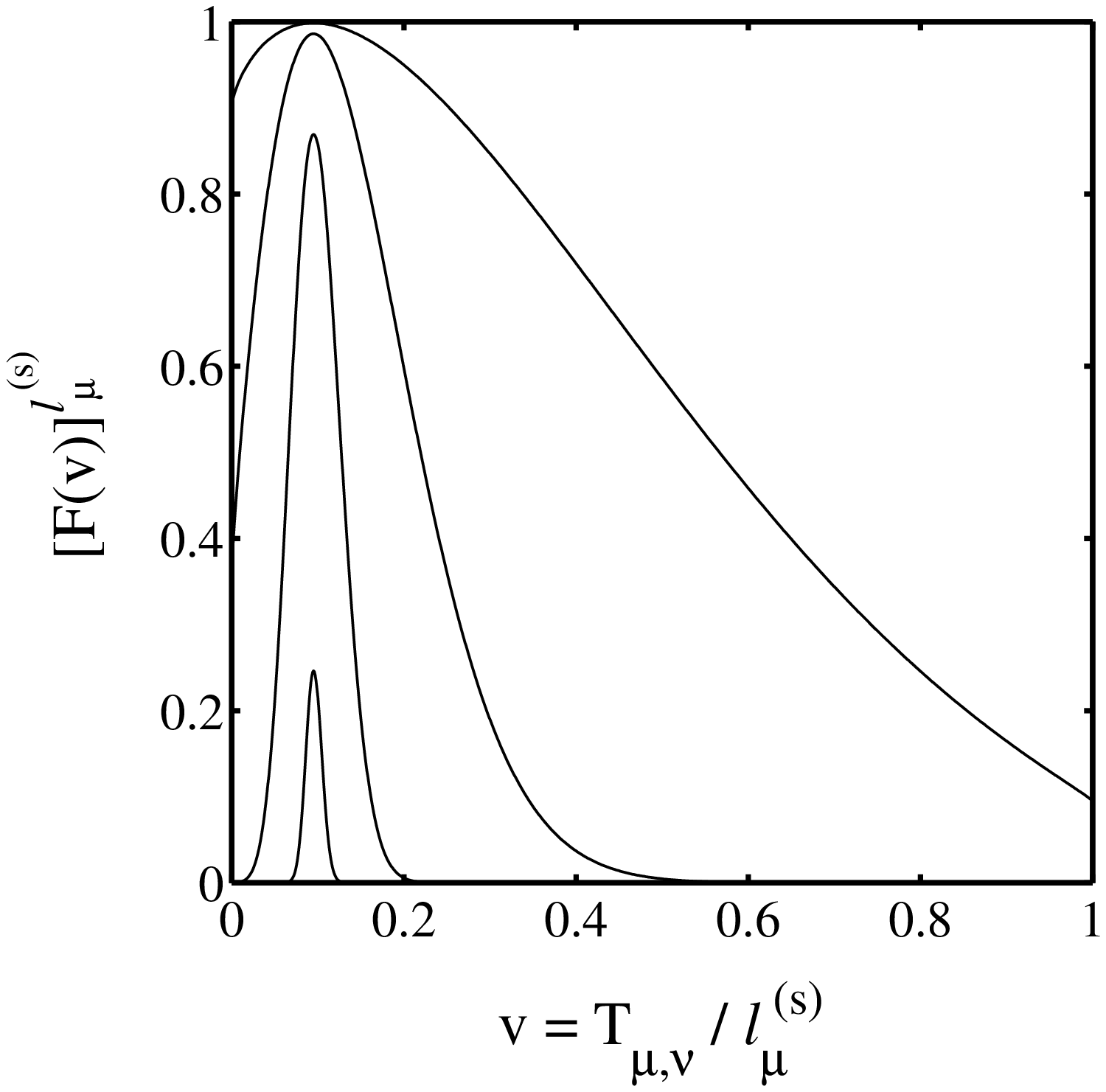,width=0.48\textwidth} \\
      \textbf{(b)} \\
      \epsfig{file=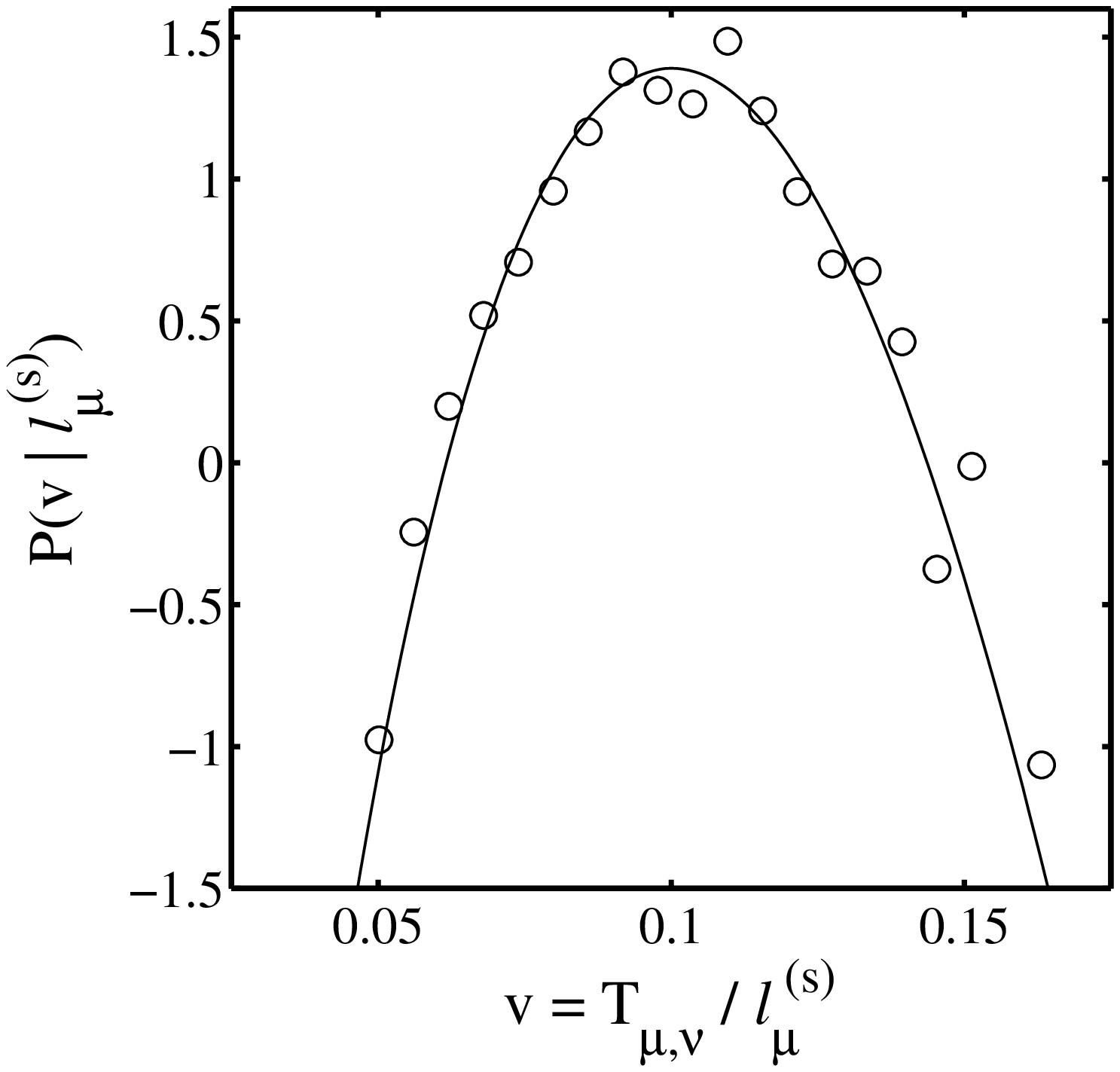,width=0.48\textwidth}
    \end{tabular}
    }
    {
    \begin{tabular}{cc}
      \textbf{(a)} & \textbf{(b)} \\
      \epsfig{file=figsche_Tell_f_theory_noname.ps,width=0.48\textwidth} &
      \epsfig{file=figsche_Tell_f_theory2_noname.ps,width=0.48\textwidth}
    \end{tabular}
    }
    \caption[Joint distribution of Tokunaga ratios and stream segment lengths]{
      Form of the joint distribution of Tokunaga ratios and stream segment lengths.
      The distribution is given in equation~\req{eq:tokunaga.probxy2}
      and is built around the function $F(v=\Tmunu/\okellsnum{\mu})$ given in
      equation~\req{eq:tokunaga.Fdef}.
      Shown in (a) is $[F(v)]^{\okellsnum{\mu}}$ for $\okellsnum{\mu}=1$, 10, 100 and 1000.
      Increasing $\okellsnum{\mu}$ corresponds to the focusing of the shape.
      In (b), the distribution $\cprob{\Tmunu}{\okellsnum{\mu}\simeq340}$
      is compared between theory (smooth curve) and data
      from the Scheidegger model (circles).  
      The Scheidegger
      model data is compiled for a range of values
      of $\okellsnum{\mu}$ rescaled as per equation~\req{eq:tokunaga.pxyrescale}.
      }
    \label{fig:tokunaga.sche_Tell_f_theory}
  \end{center}
\end{figure}

The function $F(v)=F(v;p,q)$ identified above has the form
\begin{equation}
  \label{eq:tokunaga.Fdef}
  F(v) = \rbfrac{1-v}{q}^{-(1-v)} \rbfrac{v}{p}^{-v}.
\end{equation}
where $0 < v < 1$ (here and later, the variable
$v$ will refer to $y/x$).
Note that for fixed $x$, the conditional
probability $\cprob{y}{x}$ is proportional 
to $[F(y/x)]^{x}$.
Figure~\ref{fig:tokunaga.sche_Tell_f_theory}(a)
shows $[F(v)]^x$ for 
a range of powers $x$.  
The basic function has a single peak situated near $v=p$.
For increasing $x$ which corresponds to increasing
$\okellsnum{\mu}$, the peak becomes sharper approaching
(when normalized) a delta function, i.e.,
$\lim_{x\rightarrow\infty}[F(v)]^x = \delta(v-p)$.

Figure~\ref{fig:tokunaga.sche_Tell_f_theory}(b)
provides a comparison between data for the Scheidegger
model and the analytic form of $P(\okellsnum{\mu},\Tmunu)$.
For this example, $\mu=6$ and $\nu=2$
which corresponds to $p \simeq 0.10$,
$q \simeq 0.90$ and $\tilde{p} \simeq 0.001$
(using the results of the previous section).
The smooth curve shown is 
the conditional probability $\cprob{y}{X}$ 
for the example value of $X=\okellsnum{\mu} \simeq 340$
following from equation~\req{eq:tokunaga.probxy2}.
From simulations, we obtain a discretized 
approximation to $P(\okellsnum{\mu},\Tmunu)$.
For each fixed $x=\okell$ 
in the range $165 \lesssim \okellsnum{\mu} \lesssim 345$,
we rescale the data using the following derived from
equation~\req{eq:tokunaga.probxy2},
\begin{equation}
  \label{eq:tokunaga.pxyrescale}
  P(X,y) = N' X^{-1/2}
  \left( {N'}^{-1} x^{1/2} P(x,y) \right)^{X/x}.
\end{equation}
All rescaled data is then combined, binned
and plotted as circles in
Figure~\ref{fig:tokunaga.sche_Tell_f_theory}(b),
showing excellent agreement with the
theoretical curve.

\subsection{Distributions of side branches per unit stream length}
Having obtained the general form of $P(\okellsnum{\mu}, \Tmunu)$,
we now delve further into its properties by
investigating the distributions of 
the ratio $v=\Tmunu/\okellsnum{\mu}$
and its reciprocal $w$.  

The quantity $\Tmunu/\okellsnum{\mu}$
is the number of side streams
per length of a given absorbing stream and 
when averaged over an ensemble of 
absorbing streams gives
\begin{equation}
  \label{eq:tokunaga.Tellrho}
  \avg{\Tmunu/\okellsnum{\mu}} = \rho_\nu.
\end{equation}
Accordingly, the reciprocal $\okellsnum{\mu}/\Tmunu$
is the average separation of side streams
of order $\nu$.

First, we derive $P(\Tmunu/\okell)$
from $P(\okellsnum{\mu}, \Tmunu)$.
We then consider some intuitive rescalings
which will allow us to deduce the 
form of the normalization $N(p,q)$.

We rewrite equation~\req{eq:tokunaga.probxy2} as
\begin{equation}
  \label{eq:tokunaga.probxymod}
  P(x,y) =   N' x^{-1/2}
  \exp\left\{ -x \ln \left[ -F(y/x) \right] \right\}.
\end{equation}
We transform $(x,y)$ to the modified polar coordinate system described
by $(u,v)$ with the relations
\begin{equation}
  \label{eq:tokunaga:polar}
  u^2 = x^2 + y^2
  \qquad
  \mbox{and}
  \qquad
  v = y/x.
\end{equation}
The inverse relations are $x=u/(1+v^2)$ and $y = uv/(1+v^2)$
and we also have $\dee{x}\dee{y} = x\dee{u}\dee{v}$.
Equation~\req{eq:tokunaga.probuv} leads to
\begin{equation}
  \label{eq:tokunaga.probuv}
  P(u,v) =   N' \rbfrac{u}{1+v^2}^{1/2}
  \exp\left\{ -\frac{u}{1+v^2} \ln \left[ -F(v) \right] \right\}.
\end{equation}
To find $P(v)$ we integrate out over the radial
dimension $u$:
\begin{eqnarray}
  \label{eq:tokunaga.probv}
    P(v) & = & \int_{u=0}^{\infty} \dee{u} P(u,v) \nonumber \\
    & = & N' \int_{u=0}^{\infty} \dee{u}
    \rbfrac{u}{1+v^2}^{1/2}
    \exp\left\{ -\frac{u}{1+v^2} \ln \left[ -F(v) \right] \right\}
    \nonumber \\
    & = & N' (1+v^2)(\ln[-F(v)])^{-3/2} 
    \int_{z=0}^{\infty} \dee{z} z^{1/2} e^{-z} \nonumber \\
    & = & N'' \frac{1+v^2}{(\ln[-F(v)])^{3/2}}.
\end{eqnarray}
Here, $N'' = N'\Gamma(3/2) = N'\sqrt{\pi}/2$ and
we have used the substitution
$z = u/(1+v^2)\ln[-F(v)]$.

The distribution for $w = \okellsnum{\mu}/\Tmunu = 1/v$
follows simply from equation~\req{eq:tokunaga.probv}
and we find
\begin{equation}
  \label{eq:tokunaga.probw}
  P(w) = N'' \frac{1+w^2}{w^4(\ln[-F(1/w)])^{3/2}}.
\end{equation}

\begin{figure}[tbp!]
  \begin{center}
    \ifthenelse{\boolean{@twocolumn}}
    {
    \begin{tabular}{c}
      \textbf{(a)} \\
      \epsfig{file=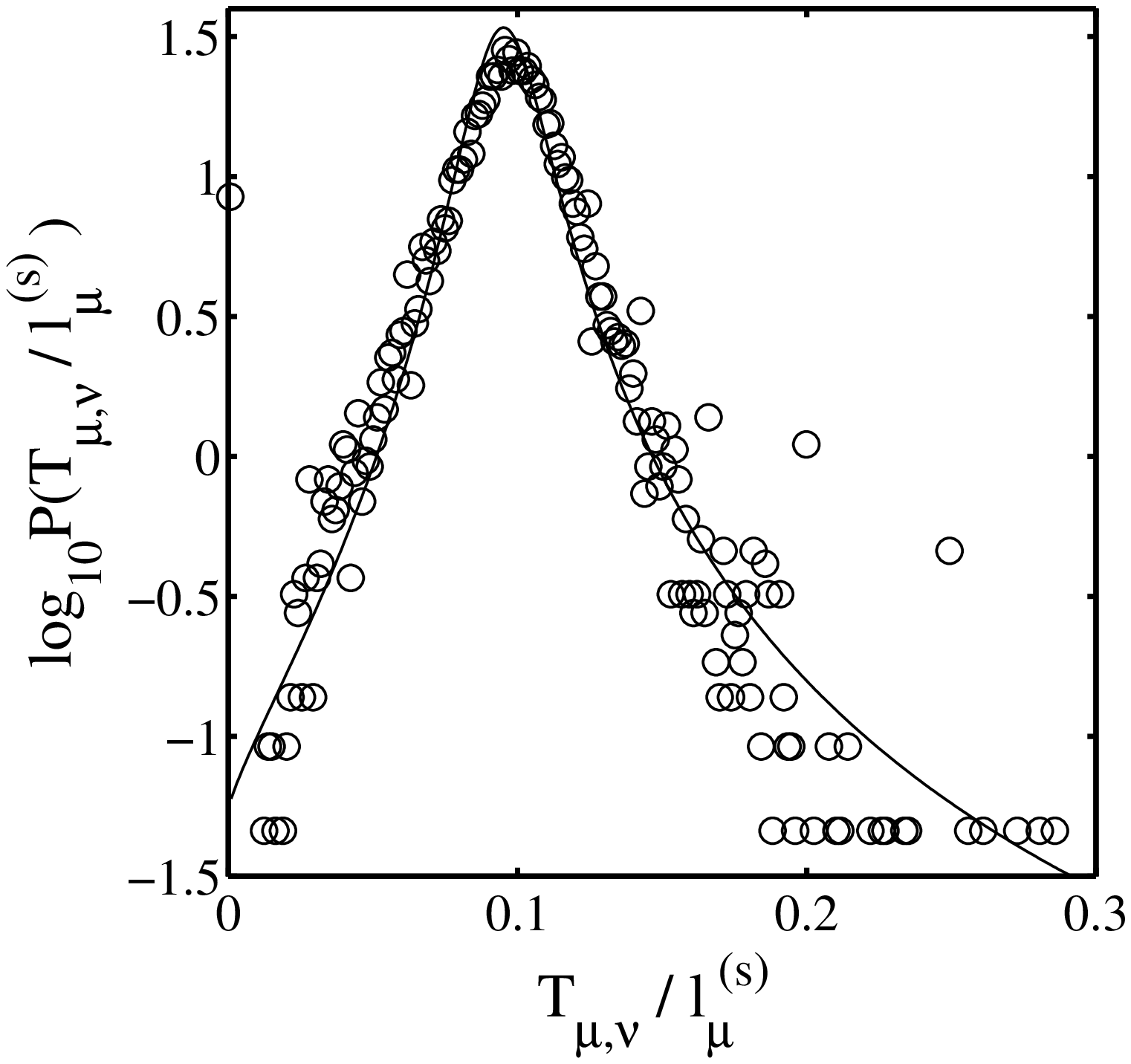,width=0.48\textwidth} \\
      \textbf{(b)} \\
      \epsfig{file=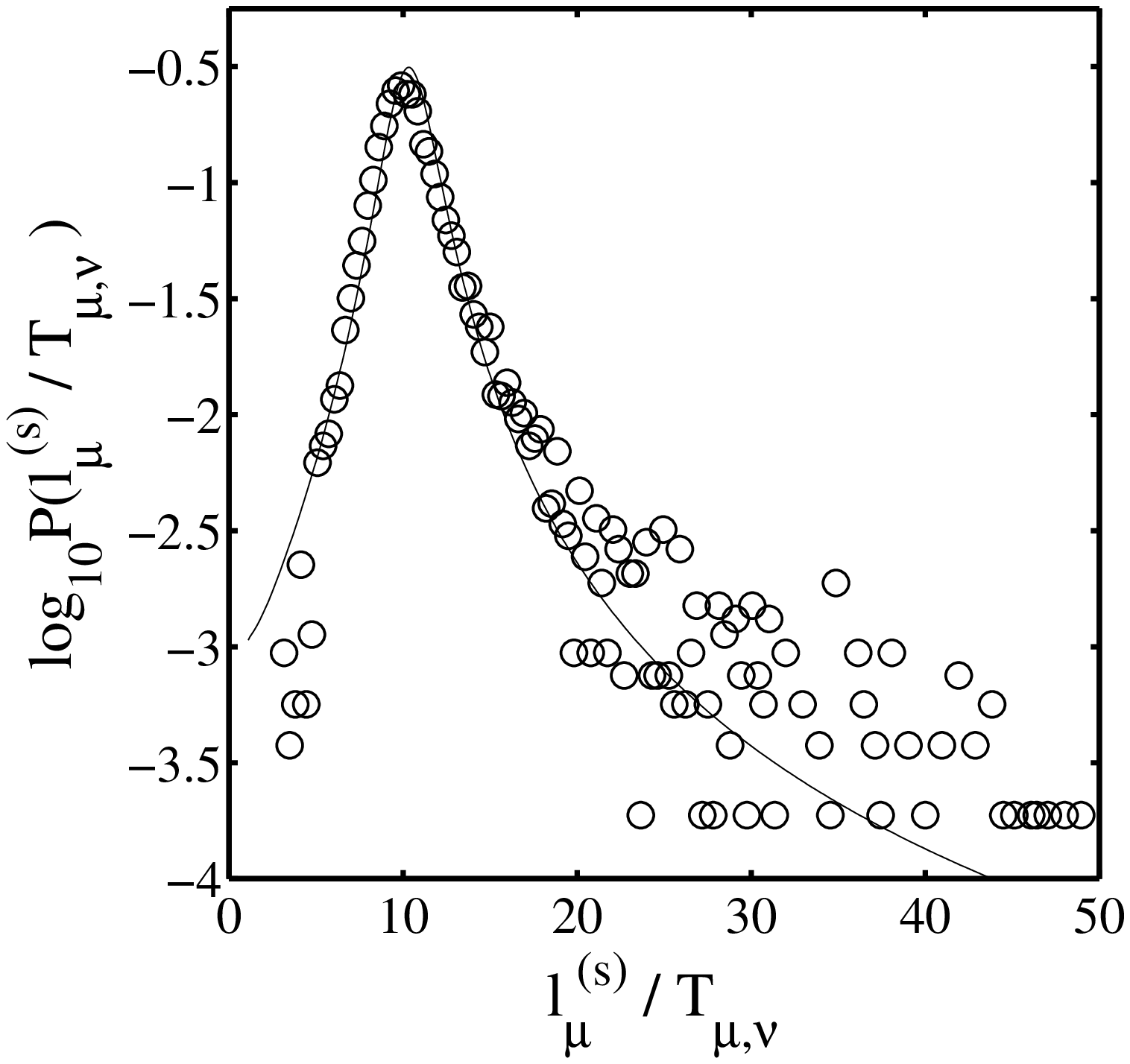,width=0.48\textwidth} 
    \end{tabular}
    }
    {
    \begin{tabular}{cc}
      \textbf{(a)} & \textbf{(b)} \\
      \epsfig{file=figsche_Tell_theory_noname.ps,width=0.48\textwidth} &
      \epsfig{file=figsche_ellT_theory_noname.ps,width=0.48\textwidth} 
    \end{tabular}
    }
    \caption[Comparison of theory with measurements of average inter-tributary spacing distribution for the Scheidegger model]{
      Comparison of theory with measurements of
      average inter-tributary distances for the Scheidegger model.
      The data in both (a) and (b) is for the case of
      order $\nu=2$ side streams and order $\mu=6$ absorbing streams.
      In (a), the distribution of $v=\Tmunu/\okellsnum{\mu}$
      obtained from the Scheidegger model (circles) is compared
      with the smooth curved predicted in equation~\req{eq:tokunaga.probv}.
      The same comparison is made for
      the reciprocal variable $w=\okellsnum{\mu}/\Tmunu$,
      the predicted curve being given in equation~\req{eq:tokunaga.probw}.
      }
    \label{fig:tokunaga.sche_ellT_theory}
  \end{center}
\end{figure}

Figures~\ref{fig:tokunaga.sche_ellT_theory}(a)
and~\ref{fig:tokunaga.sche_ellT_theory}(b)
compare the predicted forms of $P(v)$ and $P(w)$
with data from the Scheidegger model.
In both cases, the data is for order $\nu=2$
side streams being absorbed by
streams of order $\mu=6$.
Note that both distributions show an
initially exponential-like
decay away from a central peak.
Moreover, the agreement is excellent,
offering further support to the notion
that the spatial distribution
of stream segments is random.

\begin{figure}[tbp!]
  \begin{center}
    \ifthenelse{\boolean{@twocolumn}}
    {
    \begin{tabular}{c}
      \textbf{(a)} \\
      \epsfig{file=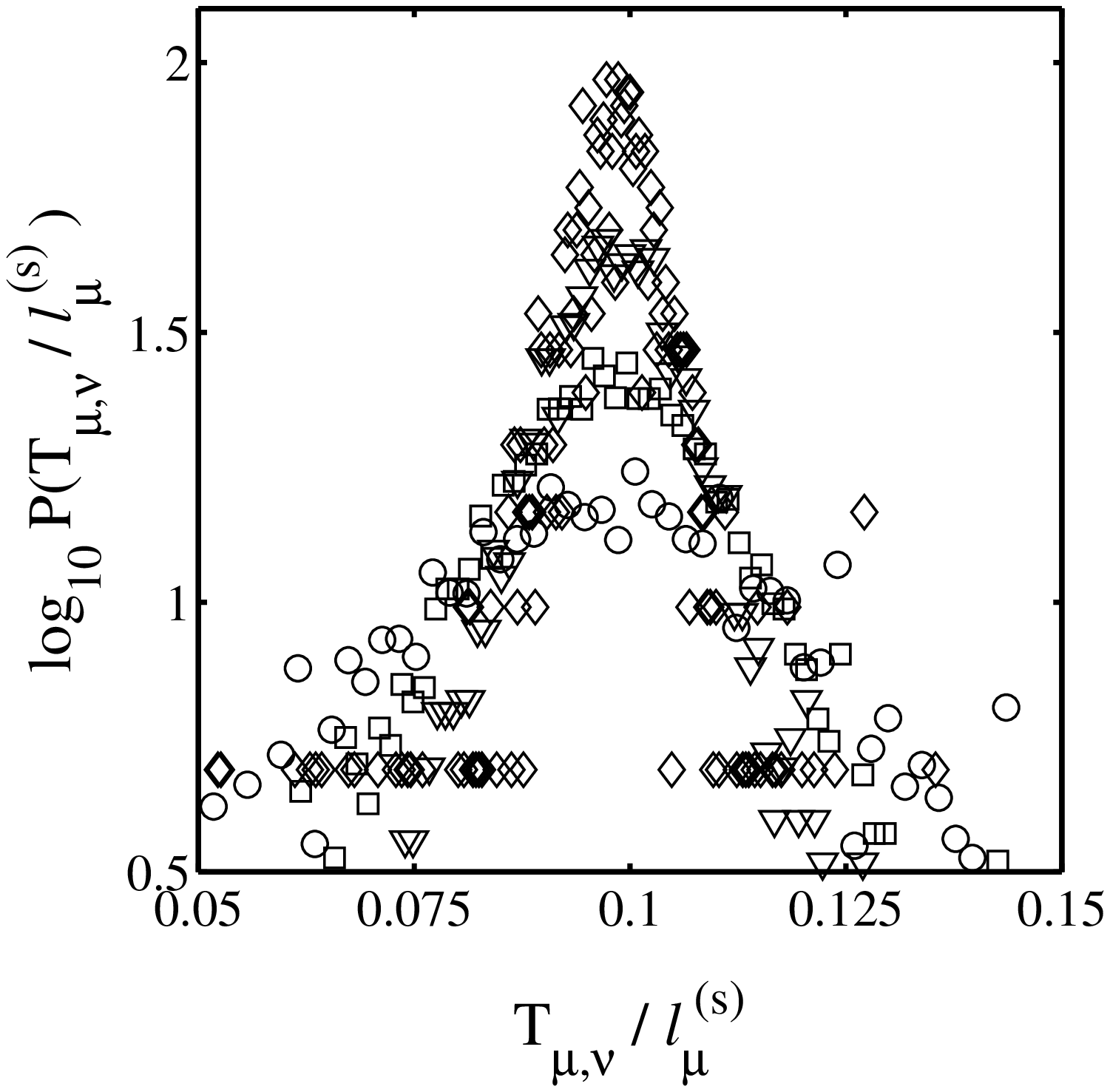,width=0.48\textwidth} \\
      \textbf{(b)} \\
      \epsfig{file=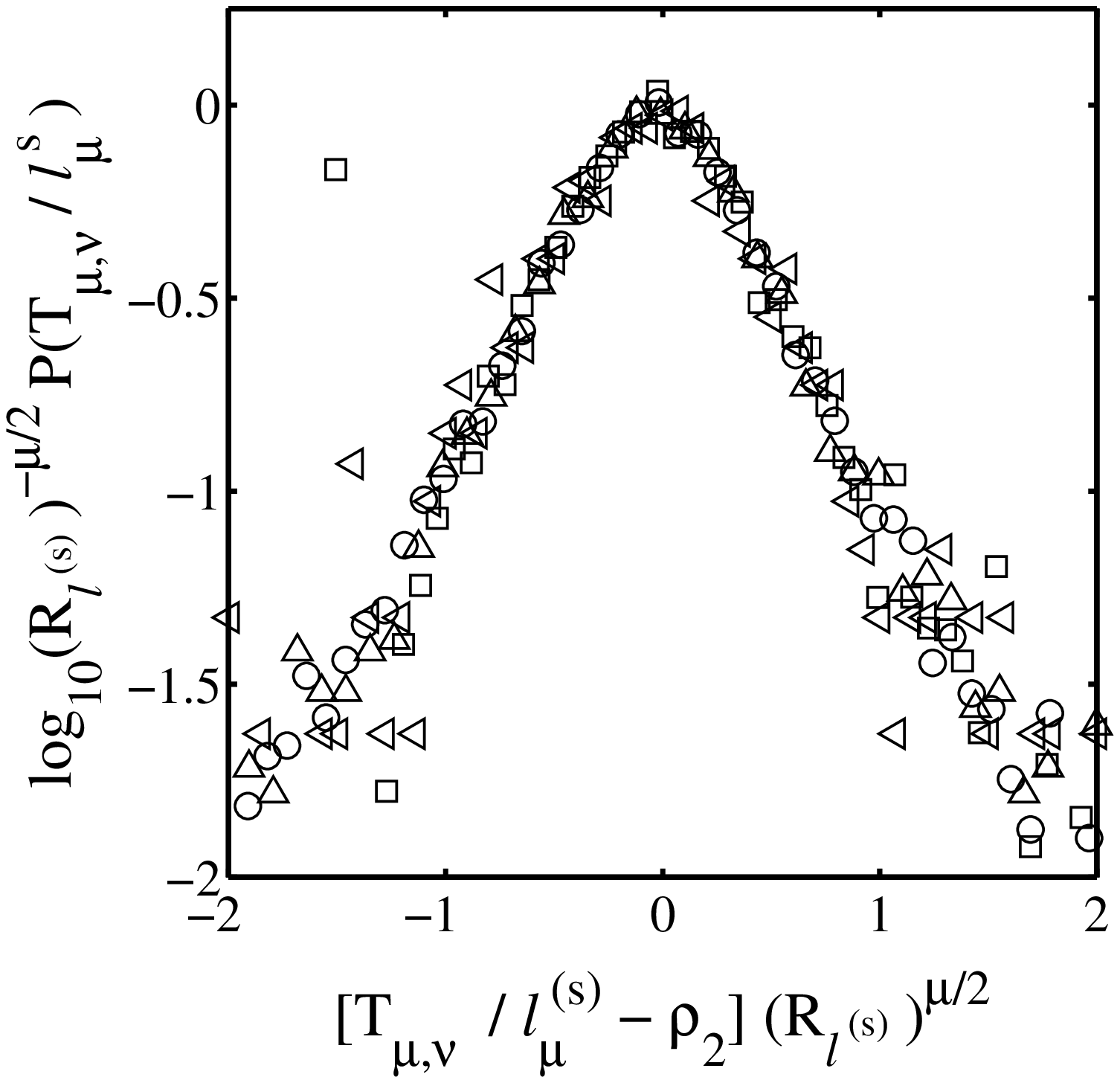,width=0.48\textwidth}
    \end{tabular}
    }
    {
    \begin{tabular}{cc}
      \textbf{(a)} & \textbf{(b)} \\
      \epsfig{file=figsche_Tell_omega_uncollapse_noname.ps,width=0.48\textwidth} &
      \epsfig{file=figsche_Tell_omega_collapse_noname.ps,width=0.48\textwidth}
    \end{tabular}
    }
    \caption[Distributions of the ratio of Tokunaga ratios to stream segment length]{
      Distributions of the quantity $\Tmunu/\okellsnum{\mu}$
      for the Scheidegger model with $\mu$, the order of the absorbing
      stream, varying and the side stream order fixed at $\nu=2$.
      Given in (a) are unrescaled distributions for 
      $\mu=5$ (circles), $\mu=6$ (squares), $\mu=7$ (triangles), and
      $\mu=8$ (diamonds).
      Note that as the order of the absorbing stream increases
      so does its typical length.  This leads to
      better averaging and the standard deviation
      of the distribution decays as $R_l^{-\om/2}$.
      The distributions are all
      centered near the typical density of order $\nu=2$ side streams, 
      $\rho_2 \simeq 0.10$.
      The rescaled versions of these distributions 
      are given in (b) with the details as per
      equation~\req{eq:tokunaga.Tellcoll1}.
      }
    \label{fig:tokunaga.sche_Tell_omega_collapse}
  \end{center}
\end{figure}

Finally, we quantify 
how changes in the orders $\mu$ and $\nu$
affect the width of the distributions
by considering some natural rescalings.
Figure~\ref{fig:tokunaga.sche_Tell_omega_collapse}(a)
shows binned, normalized distributions of
$\Tmunu/\okellsnum{\mu}$  for the Scheidegger model.  
Here, the side stream order is $\nu=2$
and the absorbing stream orders range
over $\mu=5$ to $\mu=8$.
All distributions
are centered around $\rho_2 \simeq 0.10$.

Because the average length of 
$\okellsnum{\mu}$ increases by a factor $R_\okell$ 
with $\mu$, the typical 
number of side streams increases
by the same factor.  
Since we can decompose $\okellsnum{\mu}$ as
\begin{equation}
  \label{eq:tokunaga.elldecomp}
  \okellsnum{\mu} = \okellb + \okelli + \ldots + \okelli + \okelle,
\end{equation}
where there are $\Tmunu-1$ instances of $\okelli$,
$\okellsnum{\mu}$ becomes better and better approximated
by $(\Tmunu+1)\tavg{\okelli}$.

Hence, the distribution
of $\Tmunu/\okellsnum{\mu}$ peaks up around $\rho_2$
as $\mu$ increases,
the typical width reducing by a factor of 
$1/\sqrt{R_\okell}$ for every step in $\mu$.
Using this observation,
Figure~\ref{fig:tokunaga.sche_Tell_omega_collapse}(b)
shows a rescaling of the same distributions shown in
Figure~\ref{fig:tokunaga.sche_Tell_omega_collapse}(a).
The form of this rescaling is
\begin{equation}
  \label{eq:tokunaga.Tellcoll1}
  P(\Tmunu/\okellsnum{\nu}) =
  (R_\okell)^{\mu/2} 
  G_1 \left( [\Tmunu/\okellsnum{\nu} - \rho_2](R_\okell)^{\mu/2} \right)
\end{equation}
where the function is similar to the form
of $P(v)$ given in equation~\req{eq:tokunaga.probv}.
The mean drainage density of $\rho_2$ has been subtracted
to center the distribution.  

\begin{figure}[tbp!]
  \begin{center}
    \ifthenelse{\boolean{@twocolumn}}
    {
    \begin{tabular}{c}
      \textbf{(a)} \\
      \epsfig{file=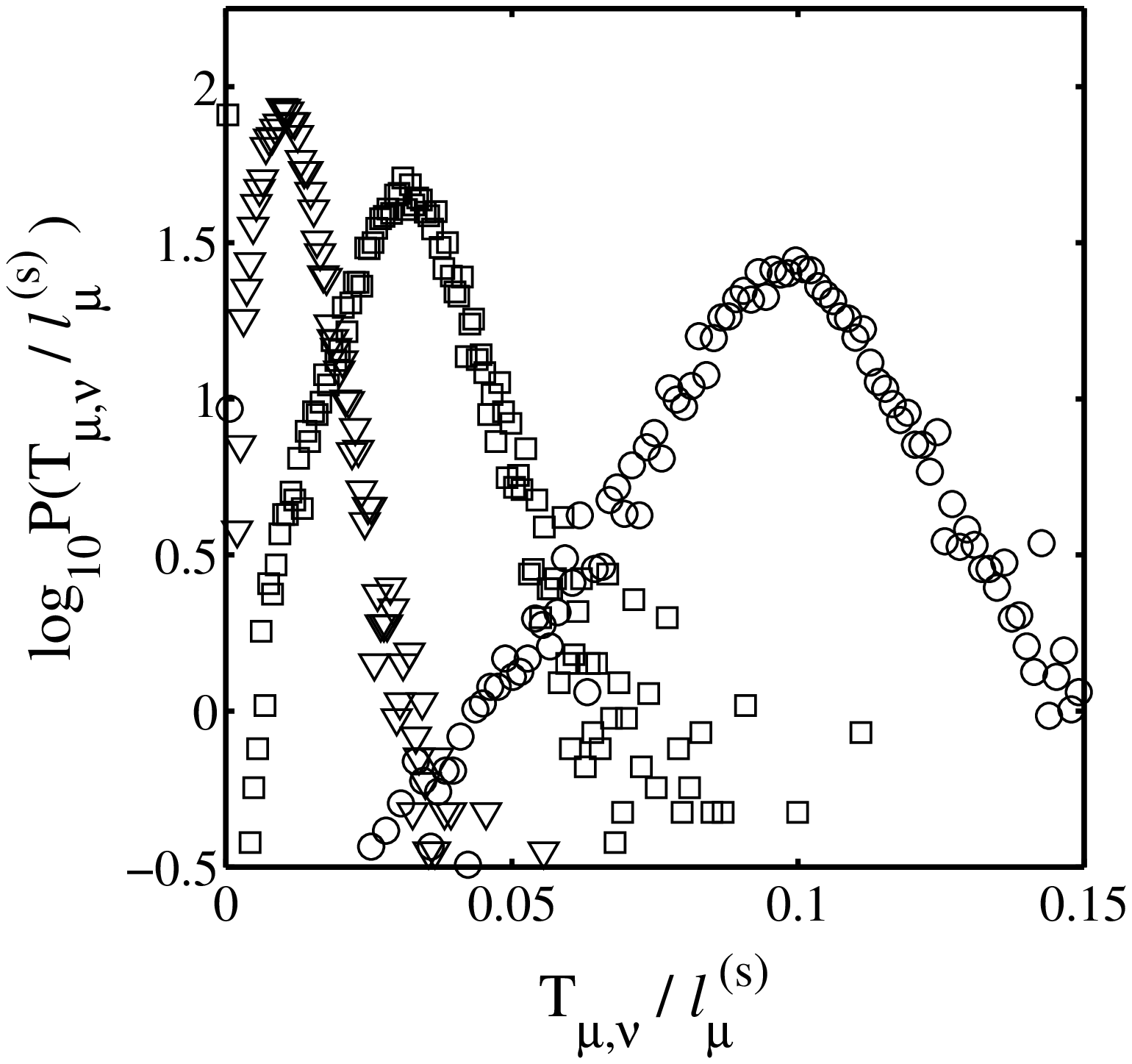,width=0.48\textwidth} \\
      \textbf{(b)} \\
      \epsfig{file=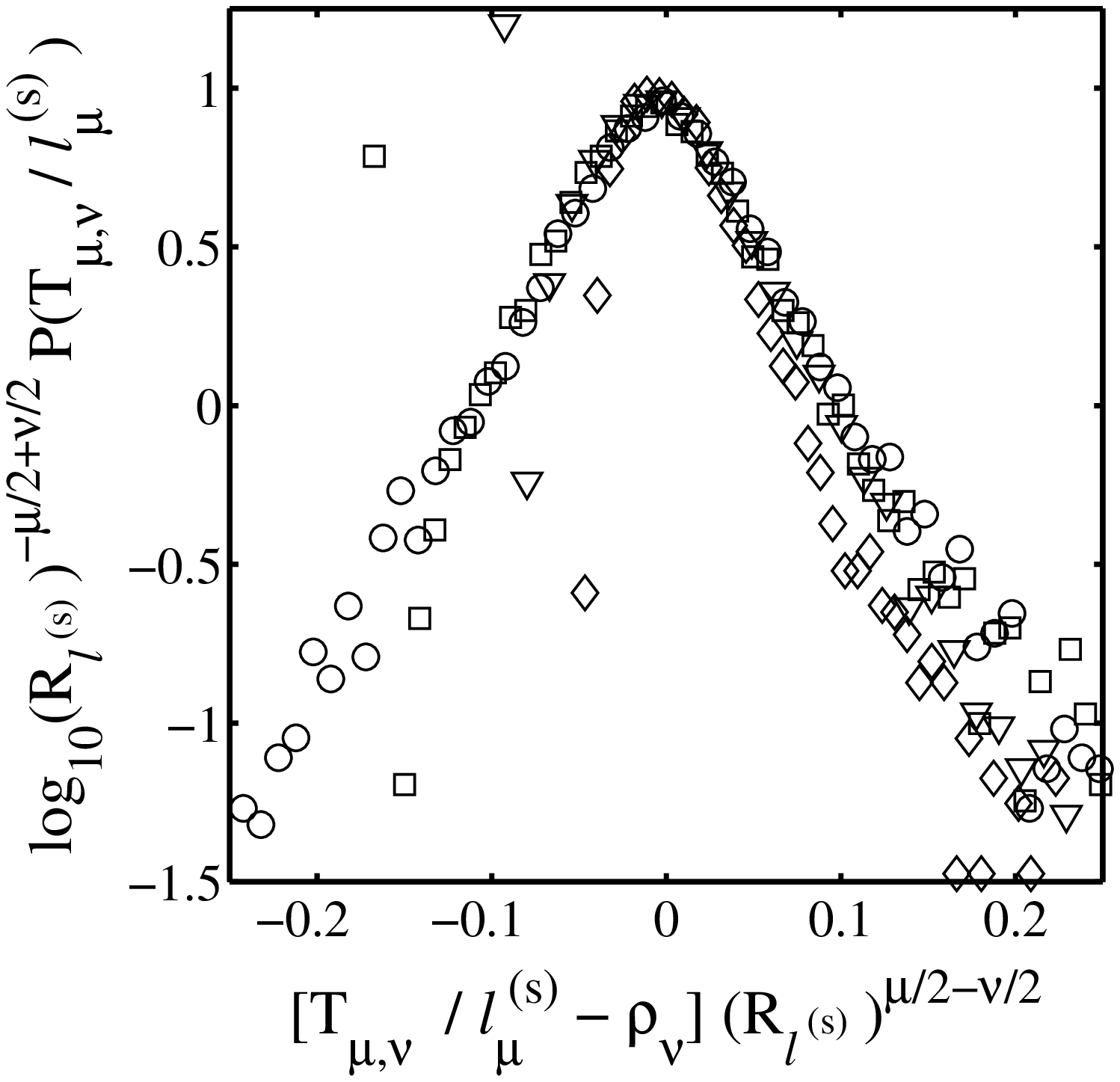,width=0.48\textwidth}
    \end{tabular}
    }
    {
    \begin{tabular}{cc}
      \textbf{(a)} & \textbf{(b)} \\
      \epsfig{file=figsche_Tell_otrib_uncollapse_noname.ps,width=0.48\textwidth} &
      \epsfig{file=figsche_Tell_otrib_collapse_noname.ps,width=0.48\textwidth}
    \end{tabular}
    }
    \caption[Distributions of number of side streams per unit length]{
      Distributions of number of side streams per unit length
      for the Scheidegger model with $\nu$, the order of side streams, varying.
      For both (a) and (b), the absorbing stream order is $\mu=6$.
      Shown in (a) are the unrescaled distributions for
      $\nu=2$ (circles), $\nu=3$ (squares),
      and $\nu=4$ (triangles).  
      Note that as $\nu$ increases, the
      mean number of side streams decreases as do the
      fluctuations.
      The distributions in (a) together with
      the distribution for $\nu=5$ (diamonds) are shown
      rescaled in (b) as per
      equation~\req{eq:tokunaga.Tellcoll3}.
      }
    \label{fig:tokunaga.sche_Tell_otrib_collapse}
  \end{center}
\end{figure}

We are able to generalize this scaling form 
of the distribution further by taking
into account side stream order.
Figures~\ref{fig:tokunaga.sche_Tell_otrib_collapse}(a)
and~\ref{fig:tokunaga.sche_Tell_otrib_collapse}(b)
respectively
show the unrescaled and rescaled distributions of $\Tmunu/\okellsnum{\mu}$
with $\nu$ allowed to vary.  This particular example
taken from the Scheidegger model
is for $\mu=6$ and the range $\nu=1$ to $\nu=5$.
Since $\nu$ is now changing,
the centers are situated at the separate values of the $\rho_\nu$.
Also, the typical number of side streams changes
with order $\nu$ so the widths of the distributions
dilate as for the varying $\mu$ case by a factor $\sqrt{R_\okell}$.
Notice that the rescaling works well for $\nu = 2,\ldots,5$
but not $\nu=1$.  As we have noted,
deviations from scaling from small orders
are to be expected.
In this case, we are led to write down
\begin{equation}
  \label{eq:tokunaga.Tellcoll2}
  P(\Tmunu/\okellsnum{\nu})
  (R_\okell)^{-\nu/2} 
  G_2\left( [\Tmunu/\okellsnum{\nu} - \rho_\nu] (R_\okell)^{-\nu/2} \right) 
\end{equation}
where, again, $G_2(z)$ is similar in form to $P(v)$.

\begin{figure}[tbp!]
  \begin{center}
    \ifthenelse{\boolean{@twocolumn}}
    {
    \begin{tabular}{c}
      \textbf{(a)} \\
      \epsfig{file=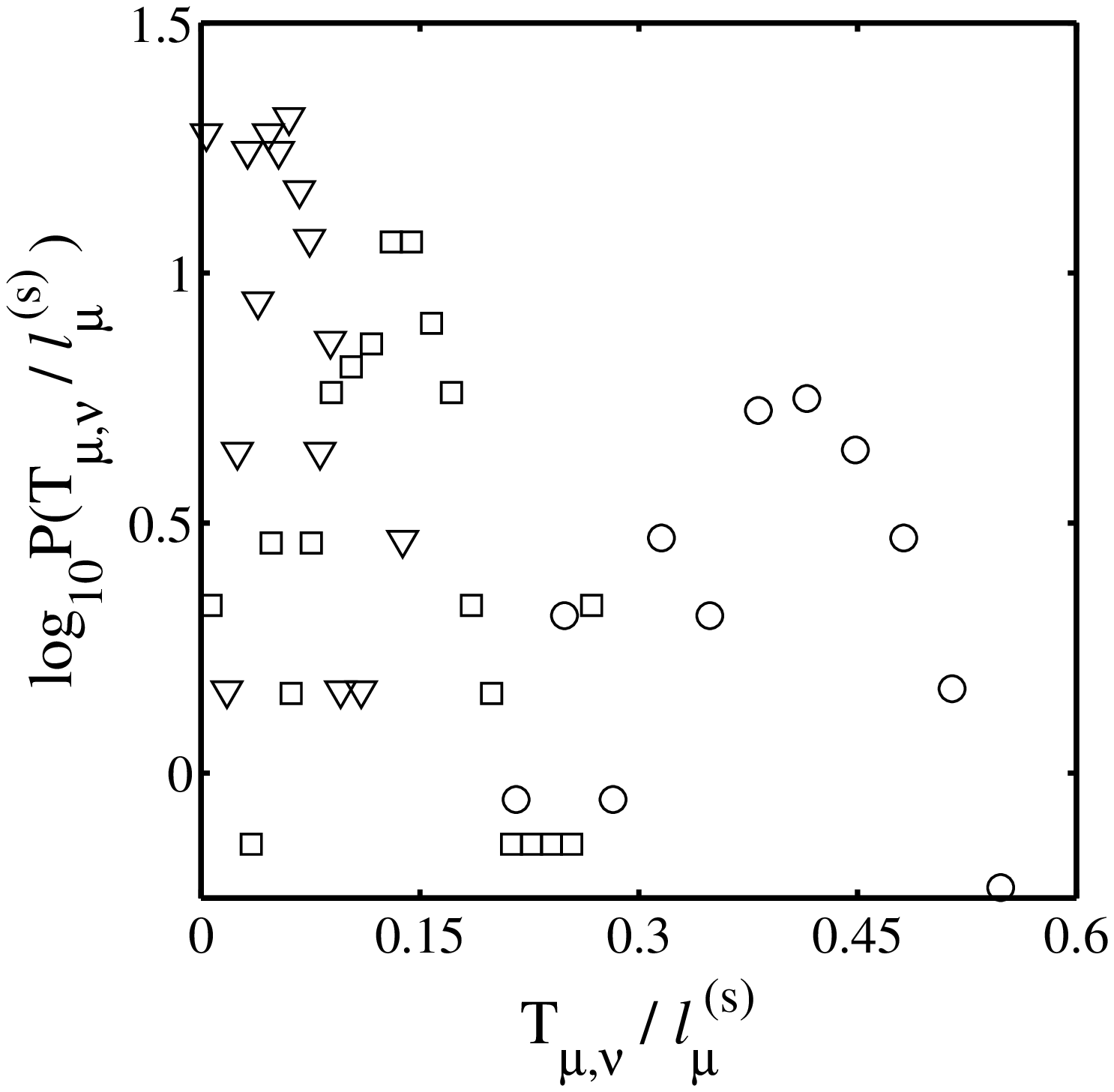,width=0.48\textwidth} \\
      \textbf{(b)} \\
      \epsfig{file=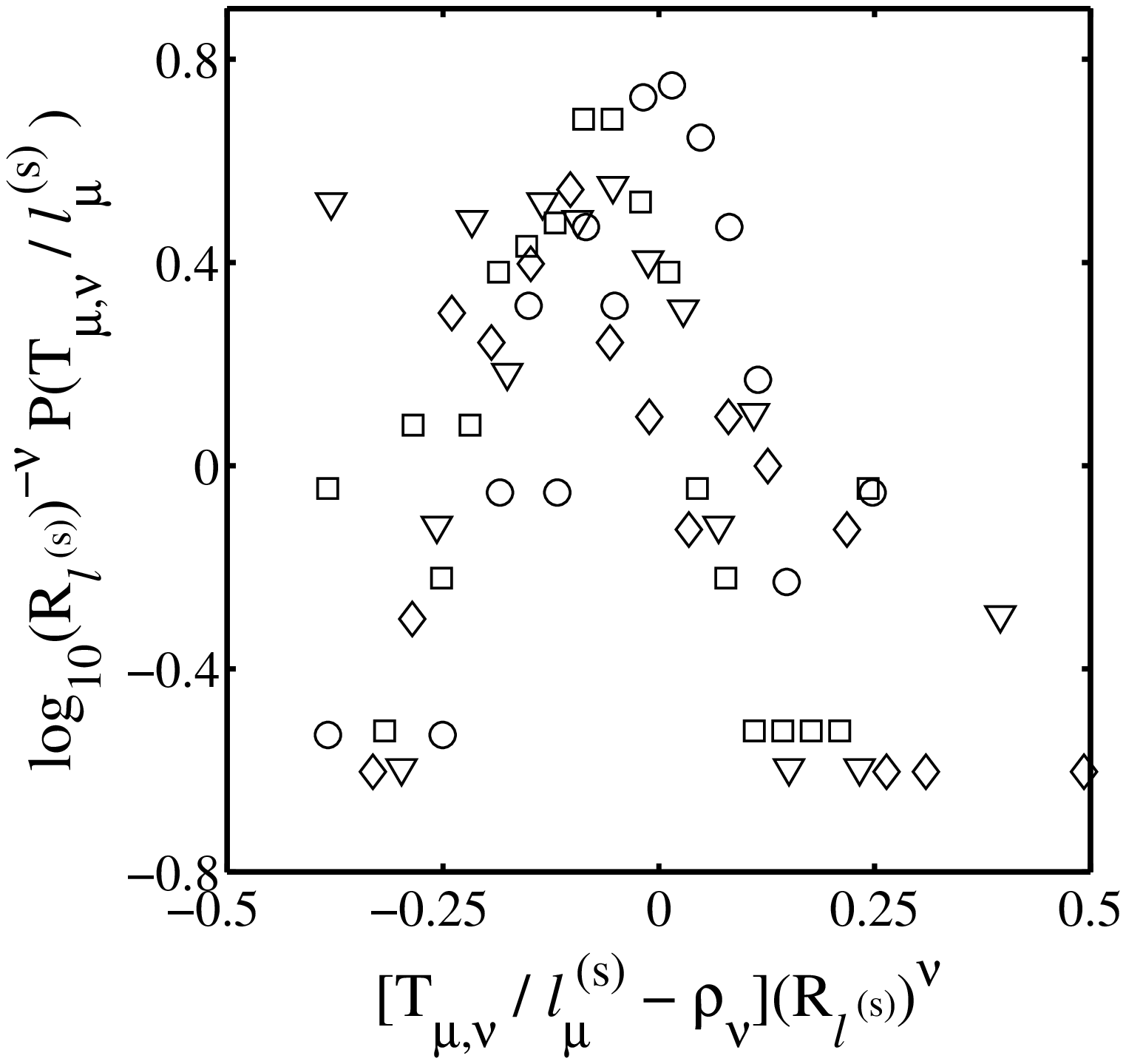,width=0.48\textwidth}
    \end{tabular}
    }
    {
    \begin{tabular}{cc}
      \textbf{(a)} & \textbf{(b)} \\
      \epsfig{file=figtokdistdata_mispi10_3_noname.ps,width=0.48\textwidth} &
      \epsfig{file=figtokdistdata_mispi10_3coll_noname.ps,width=0.48\textwidth}
    \end{tabular}
    }
    \caption[Tokunaga statistics for the Mississippi]{
      Tokunaga statistics for the Mississippi river
      basin.  The distributions are as per 
      Figure~\ref{fig:tokunaga.sche_Tell_otrib_collapse}(a),
      distributions of number of side streams per unit length
      with $\nu$, the order of side streams, varying.
      The absorbing stream order is $\mu=7$ and the
      the individual distributions correspond to
      $\nu=2$ (circles), $\nu=3$ (squares) and $\nu=4$ (triangles).
      All lengths are measured in meters.
      Rescalings of the distributions shown in (a)
      along with that for $\nu=5$ (diamonds)
      are found in (b).  Reasonable agreement
      with equation~\req{eq:tokunaga.Tellcoll3} is observed.
      }
    \label{fig:tokunaga.tokdistdata_mispi10_3}
  \end{center}
\end{figure}

We find the same rescalings apply for the Mississippi data.
For example, Figure~\ref{fig:tokunaga.tokdistdata_mispi10_3}(a)
shows unrescaled distributions of $\Tmunu/\okellsnum{\nu}$
for varying $\nu$.
Figure~\ref{fig:tokunaga.tokdistdata_mispi10_3}(b)
then shows reasonable agreement with the form
of equation~\req{eq:tokunaga.Tellcoll2}.
In this case, the Scheidegger model clearly affords
valuable guidance in our investigations of
real river networks.
The ratio $R_\okell = 2.40$ was calculated from
an analysis of $\okellsnum{\om}$ and $l_\om$.
The density $\rho_2 \simeq 0.0004$ was estimated
directly from the distributions of $\Tmunu/\okellsnum{\nu}$
and means that approximately four second-order streams appear
every ten kilometers.

Combining equations~\req{eq:tokunaga.Tellcoll3}
and~\req{eq:tokunaga.Tellcoll3}, 
we obtain the complete scaling form
\ifthenelse{\boolean{@twocolumn}}
{\begin{eqnarray}
  \label{eq:tokunaga.Tellcoll3}
  \lefteqn{P(\Tmunu/\okellsnum{\nu})} \nonumber \\
  & = & (R_\okell)^{(\mu-\nu-1)/2}
  G\left( [\Tmunu/\okellsnum{\nu}-\rho_\nu](R_\okell)^{(\mu-\nu-1)/2} \right). \nonumber\\
\end{eqnarray}}
{\begin{equation}
  \label{eq:tokunaga.Tellcoll3}
  P(\Tmunu/\okellsnum{\nu}) = (R_\okell)^{(\mu-\nu-1)/2}
  G\left( [\Tmunu/\okellsnum{\nu}-\rho_\nu](R_\okell)^{(\mu-\nu-1)/2} \right).
\end{equation}}
As per $G_1$ and $G_2$, the function $G$
is similar in form to $P(v)$.

The above scaling form makes intuitive sense
but is not obviously obtained from
an inspection of~\req{eq:tokunaga.probv}.
We therefore examine $P(v)$ by determining the
position and magnitude of its maximum.
Rather than solve $P'(v)=0$ directly, we
find an approximate solution by considering 
the argument of the denominator, $-\ln{F(v)}$,
with $F(v)$ given in equation~\req{eq:tokunaga.Fdef}.
Since the numerator of $P(v)$ is $1+v^2$ and
the maximum occurs for small $v$ this is
a justifiable step.
Setting $\tdiff{F}{v}=0$, we thus have 
\begin{equation}
  \label{eq:tokunaga.logFdiff}
  -\ln\frac{1-v}{q} + \ln{v}{p} = 0,
\end{equation}
which gives $v_m = p/(q+p) = p/(1-\tilde{p})$.
Note that for $\tilde{p} \ll 1$, 
we have $v_m \simeq p$.

Substituting $v = v_m = p/(1-\tilde{p}$
into equation~\req{eq:tokunaga.probv},
we find 
\begin{equation}
  \label{eq:tokunaga.Pvmax}
  P(v_m) \simeq N'' \tilde{p}^{-3/2} = N\tilde{p}^{-1/2}2^{-3/2}
\end{equation}
presuming $p^2 \ll 1$ and $q \simeq 1$.
Returning to the scaling form of equation~\req{eq:tokunaga.Tellcoll3},
we see that the $\tilde{p}^{-1/2}$ factor 
in equation~\req{eq:tokunaga.Pvmax} accounts for
the factors of $(R_\okell)^{\mu/2}$ since 
$\tilde{p}=\tilde{p_\mu}$ scales from level to
level by the ratio $R_\okell$.  
We therefore find the
other factor $(R_\okell)^{\nu/2}$ of
equation~\req{eq:tokunaga.Tellcoll3}
gives $N = cp^{1/2}$ where $c$ is a constant.
Since $p=p_\nu$, it is the only factor that
can provide this variation.
We thus have found the variation with 
stream order of the
normalization $N$ and have fully characterized, $P(x,y)$,
the continuum approximation of $P(\okellsnum{\mu},\Tmunu)$.

\section{Concluding remarks}
\label{sec:tokunaga.conclusion}

We have extensively investigated
river network architecture as viewed in planform.
We identify the self-similarity 
of a form of drainage density as the 
essence of the 
average connectivity and  structure of networks.  
From previous work in~\cite{dodds99pa}, 
we then understand this to be a base from which
all river network scaling laws may obtained.

We have extended the description
of tributary structure provided by Tokunaga's
law to find that side stream numbers are
distributed exponentially.  
This in turn is seen to follow from 
the fact that the length of stream segments are
themselves exponentially distributed.
We interpret this to be consequence of 
randomness in the spatial distribution
of stream segments.  Furthermore, the
presence of exponential distributions
indicate fluctuations in variables
are significant being on the order of mean values.
For the example
of stream segment lengths,
we thus identify $\xi_\okell$, 
a single parameter needed to describe all moments.
This is simply related to $\xi_t$,
which describes the distributions of Tokunaga ratios.
The exponential distribution becomes the 
null hypothesis for the distributions of these variables 
to be used in the examination of real river networks.

We are able to discern the finer details of
the connection between stream segment length
and tributary numbers.  Analysis of the 
placement of side streams along a stream
segment again reveals exponential distributions.
We are then able to postulate a joint probability
distribution for stream segment lengths
and the Tokunaga ratios.
The functional form obtained agrees well with both model
and real network data.
By further considering distributions
of the number of side streams
per unit length of individual stream segments,
we are able to capture 
how variations in the separation of
side streams are averaged out along higher-order 
absorbing streams.  
 
By expanding our knowledge of the underlying
distributions through empiricism, modeling and
theory, we obtain a more detailed picture
of network structure with which 
to compare real and theoretical networks.
We have also further shown that the simple random network model of 
Scheidegger\nocite{scheidegger67}
has an impressive ability to produce statistics whose
form may then be observed in nature.  
Indeed, the only distinction between the two is the 
exact value of the scaling exponents and ratios involved
since all distributions match up in functional form.

We end with a brief comment on
the work of Cui \etal~\cite{cui99} who have
recently also proposed a stochastic
generalization of Tokunaga's law.
They postulate that the underlying 
distribution for the $\Tmunu$ is
a negative binomial distribution.  
One parameter additional to $T_1$ and $R_T$, $\alpha$,
was introduced to reflect ``regional variability,''
i.e., statistical fluctuations in network structure.
This is in the same spirit as our
identification of a single parameter $\xi_t$.
However, our work disagrees on the nature
of the underlying distribution of $\Tmunu$.
We have consistently observed exponential distributions
for $\Tmunu$ in both model and real networks.

In closing, by finding randomness
in the spatial distribution of stream segments,
we have arrived at the most basic 
description of river network architecture.
Understanding the origin of the exact values
of quantities such as drainage density
remains an open problem.

\section*{Acknowledgements}
\label{sec:tokunaga.ack}
The authors would like to thank J.S.~Weitz for
useful discussions.
This work was supported in part by NSF grant EAR-9706220
and the Department of Energy grant DE FG02-99ER 15004.


\begin{thebibliography}{10}
\expandafter\ifx\csname bibnamefont\endcsname\relax
  \def\bibnamefont#1{#1}\fi
\expandafter\ifx\csname bibfnamefont\endcsname\relax
  \def\bibfnamefont#1{#1}\fi
\expandafter\ifx\csname url\endcsname\relax
  \def\url#1{\texttt{#1}}\fi
\expandafter\ifx\csname urlprefix\endcsname\relax\def\urlprefix{URL }\fi
\expandafter\ifx\csname bibinfo\endcsname\relax \def\bibinfo#1#2{#2}\fi
\expandafter\ifx\csname eprint\endcsname\relax \def\eprint#1{#1}\fi

\bibitem{dodds2000ua}
\bibinfo{author}{\bibfnamefont{P.~S.} \bibnamefont{Dodds}} \bibnamefont{and}
  \bibinfo{author}{\bibfnamefont{D.~H.} \bibnamefont{Rothman}},
  \emph{\bibinfo{title}{Geometry of River Networks {I}: {S}caling,
  Fluctuations, and Deviations}} (\bibinfo{year}{2000}),
  \bibinfo{note}{submitted to PRE}.

\bibitem{maritan96a}
\bibinfo{author}{\bibfnamefont{A.}~\bibnamefont{Maritan}},
  \bibinfo{author}{\bibfnamefont{A.}~\bibnamefont{Rinaldo}},
  \bibinfo{author}{\bibfnamefont{R.}~\bibnamefont{Rigon}},
  \bibinfo{author}{\bibfnamefont{A.}~\bibnamefont{Giacometti}},
  \bibnamefont{and}
  \bibinfo{author}{\bibfnamefont{I.}~\bibnamefont{Rodr\'{\i}guez-Iturbe}},
  \bibinfo{journal}{Phys. Rev. E}
  \textbf{\bibinfo{volume}{53}}(\bibinfo{number}{2}), \bibinfo{pages}{1510}
  (\bibinfo{year}{1996}).

\bibitem{rodriguez-iturbe97}
\bibinfo{author}{\bibfnamefont{I.}~\bibnamefont{Rodr\'{\i}guez-Iturbe}}
  \bibnamefont{and} \bibinfo{author}{\bibfnamefont{A.}~\bibnamefont{Rinaldo}},
  \emph{\bibinfo{title}{Fractal River Basins: Chance and Self-Organization}}
  (\bibinfo{publisher}{Cambridge University Press}, \bibinfo{address}{Great
  Britain}, \bibinfo{year}{1997}).

\bibitem{dodds99pa}
\bibinfo{author}{\bibfnamefont{P.~S.} \bibnamefont{Dodds}} \bibnamefont{and}
  \bibinfo{author}{\bibfnamefont{D.~H.} \bibnamefont{Rothman}},
  \bibinfo{journal}{Phys. Rev. E}
  \textbf{\bibinfo{volume}{59}}(\bibinfo{number}{5}), \bibinfo{pages}{4865}
  (\bibinfo{year}{1999}), \eprint{cond-mat/9808244}.

\bibitem{dodds2000pa}
\bibinfo{author}{\bibfnamefont{P.~S.} \bibnamefont{Dodds}} \bibnamefont{and}
  \bibinfo{author}{\bibfnamefont{D.~H.} \bibnamefont{Rothman}},
  \bibinfo{journal}{Annu. Rev. Earth Planet. Sci.}
  \textbf{\bibinfo{volume}{28}}, \bibinfo{pages}{571} (\bibinfo{year}{2000}).

\bibitem{dodds2000ub}
\bibinfo{author}{\bibfnamefont{P.~S.} \bibnamefont{Dodds}} \bibnamefont{and}
  \bibinfo{author}{\bibfnamefont{D.~H.} \bibnamefont{Rothman}},
  \emph{\bibinfo{title}{Geometry of River Networks {I}{I}: {D}istributions of
  Component Size and Number}} (\bibinfo{year}{2000}), \bibinfo{note}{submitted
  to PRE}.

\bibitem{tokunaga66}
\bibinfo{author}{\bibfnamefont{E.}~\bibnamefont{Tokunaga}},
  \bibinfo{journal}{Geophys. Bull. Hokkaido Univ.}
  \textbf{\bibinfo{volume}{15}}, \bibinfo{pages}{1} (\bibinfo{year}{1966}).

\bibitem{tokunaga78}
\bibinfo{author}{\bibfnamefont{E.}~\bibnamefont{Tokunaga}},
  \bibinfo{journal}{Geogr. Rep., Tokyo Metrop. Univ.}
  \textbf{\bibinfo{volume}{13}}, \bibinfo{pages}{1} (\bibinfo{year}{1978}).

\bibitem{tokunaga84}
\bibinfo{author}{\bibfnamefont{E.}~\bibnamefont{Tokunaga}},
  \bibinfo{journal}{Trans. Jpn. Geomorphol. Union}
  \textbf{\bibinfo{volume}{5}}(\bibinfo{number}{2}), \bibinfo{pages}{71}
  (\bibinfo{year}{1984}).

\bibitem{scheidegger67}
\bibinfo{author}{\bibfnamefont{A.~E.} \bibnamefont{Scheidegger}},
  \bibinfo{journal}{Bull. Int. Assoc. Sci. Hydrol.}
  \textbf{\bibinfo{volume}{12}}(\bibinfo{number}{1}), \bibinfo{pages}{15}
  (\bibinfo{year}{1967}).

\bibitem{horton45}
\bibinfo{author}{\bibfnamefont{R.~E.} \bibnamefont{Horton}},
  \bibinfo{journal}{Bull. Geol. Soc. Am}
  \textbf{\bibinfo{volume}{56}}(\bibinfo{number}{3}), \bibinfo{pages}{275}
  (\bibinfo{year}{1945}).

\bibitem{strahler57}
\bibinfo{author}{\bibfnamefont{A.~N.} \bibnamefont{Strahler}},
  \bibinfo{journal}{EOS Trans. AGU}
  \textbf{\bibinfo{volume}{38}}(\bibinfo{number}{6}), \bibinfo{pages}{913}
  (\bibinfo{year}{1957}).

\bibitem{cui99}
\bibinfo{author}{\bibfnamefont{G.}~\bibnamefont{Cui}},
  \bibinfo{author}{\bibfnamefont{B.}~\bibnamefont{Williams}}, \bibnamefont{and}
  \bibinfo{author}{\bibfnamefont{G.}~\bibnamefont{Kuczera}},
  \bibinfo{journal}{Water Resour. Res.}
  \textbf{\bibinfo{volume}{35}}(\bibinfo{number}{10}), \bibinfo{pages}{3139}
  (\bibinfo{year}{1999}).

\bibitem{turcotte98}
\bibinfo{author}{\bibfnamefont{D.~L.} \bibnamefont{Turcotte}},
  \bibinfo{author}{\bibfnamefont{J.~D.} \bibnamefont{Pelletier}},
  \bibnamefont{and} \bibinfo{author}{\bibfnamefont{W.~I.}
  \bibnamefont{Newman}}, \bibinfo{journal}{J. Theor. Biol.}
  \textbf{\bibinfo{volume}{193}}, \bibinfo{pages}{577} (\bibinfo{year}{1998}).

\bibitem{peckham95}
\bibinfo{author}{\bibfnamefont{S.~D.} \bibnamefont{Peckham}},
  \bibinfo{journal}{Water Resour. Res.}
  \textbf{\bibinfo{volume}{31}}(\bibinfo{number}{4}), \bibinfo{pages}{1023}
  (\bibinfo{year}{1995}).

\bibitem{schumm56a}
\bibinfo{author}{\bibfnamefont{S.~A.} \bibnamefont{Schumm}},
  \bibinfo{journal}{Bull. Geol. Soc. Am} \textbf{\bibinfo{volume}{67}},
  \bibinfo{pages}{597} (\bibinfo{year}{1956}).

\bibitem{huxley36}
\bibinfo{author}{\bibfnamefont{J.~S.} \bibnamefont{Huxley}} \bibnamefont{and}
  \bibinfo{author}{\bibfnamefont{G.}~\bibnamefont{Teissier}},
  \bibinfo{journal}{Nature} \textbf{\bibinfo{volume}{137}},
  \bibinfo{pages}{780} (\bibinfo{year}{1936}).

\bibitem{hack57}
\bibinfo{author}{\bibfnamefont{J.~T.} \bibnamefont{Hack}},
  \bibinfo{journal}{U.S. Geol. Surv. Prof. Pap.}
  \textbf{\bibinfo{volume}{294-B}}, \bibinfo{pages}{45} (\bibinfo{year}{1957}).

\bibitem{colaiori97}
\bibinfo{author}{\bibfnamefont{F.}~\bibnamefont{Colaiori}},
  \bibinfo{author}{\bibfnamefont{A.}~\bibnamefont{Flammini}},
  \bibinfo{author}{\bibfnamefont{A.}~\bibnamefont{Maritan}}, \bibnamefont{and}
  \bibinfo{author}{\bibfnamefont{J.~R.} \bibnamefont{Banavar}},
  \bibinfo{journal}{Phys. Rev. E}
  \textbf{\bibinfo{volume}{55}}(\bibinfo{number}{2}), \bibinfo{pages}{1298}
  (\bibinfo{year}{1997}).

\bibitem{tarboton90}
\bibinfo{author}{\bibfnamefont{D.~G.} \bibnamefont{Tarboton}},
  \bibinfo{author}{\bibfnamefont{R.~L.} \bibnamefont{Bras}}, \bibnamefont{and}
  \bibinfo{author}{\bibfnamefont{I.}~\bibnamefont{Rodr\'{\i}guez-Iturbe}},
  \bibinfo{journal}{Water Resour. Res.}
  \textbf{\bibinfo{volume}{26}}(\bibinfo{number}{9}), \bibinfo{pages}{2243}
  (\bibinfo{year}{1990}).

\bibitem{manna92}
\bibinfo{author}{\bibfnamefont{S.~S.} \bibnamefont{Manna}},
  \bibinfo{author}{\bibfnamefont{D.}~\bibnamefont{Dhar}}, \bibnamefont{and}
  \bibinfo{author}{\bibfnamefont{S.~N.} \bibnamefont{Majumdar}},
  \bibinfo{journal}{Phys. Rev. A} \textbf{\bibinfo{volume}{46}},
  \bibinfo{pages}{4471} (\bibinfo{year}{1992}).

\bibitem{manna96}
\bibinfo{author}{\bibfnamefont{S.~S.} \bibnamefont{Manna}} \bibnamefont{and}
  \bibinfo{author}{\bibfnamefont{B.}~\bibnamefont{Subramanian}},
  \bibinfo{journal}{Phys. Rev. Lett.}
  \textbf{\bibinfo{volume}{76}}(\bibinfo{number}{18}), \bibinfo{pages}{3460}
  (\bibinfo{year}{1996}).

\bibitem{feller68I}
\bibinfo{author}{\bibfnamefont{W.}~\bibnamefont{Feller}},
  \emph{\bibinfo{title}{An Introduction to Probability Theory and Its
  Applications}}, vol.~\bibinfo{volume}{I} (\bibinfo{publisher}{John Wiley \&
  Sons}, \bibinfo{address}{New York}, \bibinfo{year}{1968}), third ed.

\bibitem{takayasu88}
\bibinfo{author}{\bibfnamefont{H.}~\bibnamefont{Takayasu}},
  \bibinfo{author}{\bibfnamefont{I.}~\bibnamefont{Nishikawa}},
  \bibnamefont{and} \bibinfo{author}{\bibfnamefont{H.}~\bibnamefont{Tasaki}},
  \bibinfo{journal}{Phys. Rev. A}
  \textbf{\bibinfo{volume}{37}}(\bibinfo{number}{8}), \bibinfo{pages}{3110}
  (\bibinfo{year}{1988}).

\bibitem{takayasu89a}
\bibinfo{author}{\bibfnamefont{M.}~\bibnamefont{Takayasu}} \bibnamefont{and}
  \bibinfo{author}{\bibfnamefont{H.}~\bibnamefont{Takayasu}},
  \bibinfo{journal}{Phys. Rev. A}
  \textbf{\bibinfo{volume}{39}}(\bibinfo{number}{8}), \bibinfo{pages}{4345}
  (\bibinfo{year}{1989}).

\bibitem{takayasu91}
\bibinfo{author}{\bibfnamefont{H.}~\bibnamefont{Takayasu}},
  \bibinfo{author}{\bibfnamefont{M.}~\bibnamefont{Takayasu}},
  \bibinfo{author}{\bibfnamefont{A.}~\bibnamefont{Provata}}, \bibnamefont{and}
  \bibinfo{author}{\bibfnamefont{G.}~\bibnamefont{Huber}}, \bibinfo{journal}{J.
  Stat. Phys.} \textbf{\bibinfo{volume}{65}}(\bibinfo{number}{3/4}),
  \bibinfo{pages}{725} (\bibinfo{year}{1991}).

\bibitem{huber91}
\bibinfo{author}{\bibfnamefont{G.}~\bibnamefont{Huber}},
  \bibinfo{journal}{Physica A} \textbf{\bibinfo{volume}{170}},
  \bibinfo{pages}{463} (\bibinfo{year}{1991}).

\bibitem{dhar99}
\bibinfo{author}{\bibfnamefont{D.}~\bibnamefont{Dhar}},
  \bibinfo{journal}{Physica A} \textbf{\bibinfo{volume}{263}},
  \bibinfo{pages}{4} (\bibinfo{year}{1999}).

\bibitem{bender78}
\bibinfo{author}{\bibfnamefont{C.~M.} \bibnamefont{Bender}} \bibnamefont{and}
  \bibinfo{author}{\bibfnamefont{S.~A.} \bibnamefont{Orszag}},
  \emph{\bibinfo{title}{Advanced mathematical methods for scientists and
  engineers}}, International series in pure and applied mathematics
  (\bibinfo{publisher}{McGraw-Hill}, \bibinfo{address}{New York},
  \bibinfo{year}{1978}).

\end{thebibliography}
\end{document}